\documentclass[11pt]{article}
\pdfoutput=1
\usepackage[linktocpage=true]{hyperref}
\usepackage{color}

\usepackage{graphicx}
\usepackage[textwidth = 430 pt, textheight = 630 pt]{geometry}

\definecolor{MyDarkBlue}{rgb}{0.15,0.25,0.45}
\usepackage{epsfig,rotating}
\usepackage{amsmath,amssymb}
\usepackage{amsfonts}
\usepackage{mathrsfs}
\usepackage{bbm}

\usepackage{latexsym}
\usepackage{amsthm}

\usepackage{stmaryrd}

\usepackage[utf8x]{inputenc}

\usepackage{hyperref}
\hypersetup{
colorlinks=true,
citecolor=MyDarkBlue,
linkcolor=MyDarkBlue,
urlcolor=MyDarkBlue,
pdfauthor={Sam Palmer and Christian S\"amann},
pdftitle={Six-Dimensional (1,0) Superconformal Models and Higher Gauge Theory},
pdfsubject={hep-th}
breaklinks=true
}

\usepackage{tikz}
\usepackage{mathtools}

\theoremstyle{remark}

\theoremstyle{definition}

\theoremstyle{remark}

\let\fn\footnote
\renewcommand{\footnote}[1]{\linespread{1.1}\fn{#1}\linespread{1.29}}

\makeatletter\renewcommand{\section}{\@startsection
{section}{1}{\z@}{-3.5ex plus -1ex minus
    -.2ex}{2.3ex plus .2ex}{\bf }}
\makeatletter\renewcommand{\subsection}{\@startsection{subsection}{2}{\z@}{-3.25ex
plus -1ex minus
   -.2ex}{1.5ex plus .2ex}{\it }}
\makeatletter\renewcommand{\subsubsection}{\@startsection{subsubsection}{3}{-2.45ex}{-3.25ex
plus -1ex minus -.2ex}{1.5ex plus .2ex}{\it }}
\renewcommand{\thesection}{\arabic{section}}
\renewcommand{\thesubsection}{\arabic{section}.\arabic{subsection}}
\renewcommand{\@seccntformat}[1]{\@nameuse{the#1}.~~}

\renewcommand{\theequation}{\thesection.\arabic{equation}}
\makeatletter \@addtoreset{equation}{section}

\newcommand{\appendices}{
\section*{Appendix}\label{appendices}\setcounter{subsection}{0}
\addcontentsline{toc}{section}{Appendix}
\setcounter{equation}{0}
\makeatletter
\renewcommand{\theequation}{\Alph{subsection}.\arabic{equation}}
\renewcommand{\thesubsection}{\Alph{subsection}}
\@addtoreset{equation}{subsection}
\makeatother
}

\def\slasha#1{\setbox0=\hbox{$#1$}#1\hskip-\wd0\hbox to\wd0{\hss\sl/\/\hss}}

\def\periodb#1{\setbox0=\hbox{$#1$}#1\hskip-\wd0\hbox to\wd0{-}}

\newcommand{\im}{\mathrm{im}}   			

\newcommand{\CA}{\mathcal{A}}    			

\newcommand{\CC}{\mathcal{C}}

\newcommand{\CD}{\mathcal{D}}

\newcommand{\CF}{\mathcal{F}}

\newcommand{\CH}{\mathcal{H}}

\newcommand{\CN}{\mathcal{N}}

\newcommand{\CR}{\mathcal{R}}

\newcommand{\CE}{\mathcal{E}}
\newcommand{\CEa}{\mathsf{CE}}
\newcommand{\fra}{\mathfrak{a}}				
\newcommand{\frb}{\mathfrak{b}}				
\newcommand{\frcs}{\mathfrak{cs}}				
\newcommand{\frg}{\mathfrak{g}}				
\newcommand{\frh}{\mathfrak{h}}				
\newcommand{\frl}{\mathfrak{l}}				

\newcommand{\fru}{\mathfrak{u}}

\newcommand{\sft}{{\sf t}}
\newcommand{\sfm}{{\sf m}}
\newcommand{\sfg}{{\sf g}}
\newcommand{\sfh}{{\sf h}}
\newcommand{\sfd}{{\sf d}}
\newcommand{\sfb}{{\sf b}}
\newcommand{\sff}{{\sf f}}
\newcommand{\sfk}{{\sf k}}

\newcommand{\FR}{\mathbbm{R}}     			
\newcommand{\RZ}{\mathbbm{Z}}     			

\newcommand{\lambdab}{\bar{\lambda}}

\newcommand{\dd}{\mathrm{d}}     			
\newcommand{\dpar}{\partial}     			
\newcommand{\chib}{{\bar{\chi}}}   	  		
\newcommand{\eps}{{\varepsilon}}			
\newcommand{\epsb}{{\bar{\varepsilon}}}			

\newcommand{\eand}{{\qquad\mbox{and}\qquad}}     		

\newcommand{\kernel}{{\mathrm{ker}}}

\newcommand{\au}{\mathfrak{u}}

\newcommand{\aso}{\mathfrak{so}}

\newcommand{\sG}{\mathsf{G}}

\newcommand{\sEnd}{\mathsf{End}\,}

\newcommand*{\longhookrightarrow}{\ensuremath{\lhook\joinrel\relbar\joinrel\rightarrow}}

\newcommand{\acton}{\vartriangleright}     			
\def\tyng(#1){\hbox{\tiny$\yng(#1)$}}			
\def\tyoung(#1){\hbox{\tiny$\young(#1)$}}			

\newcommand{\beq}{\begin{eqnarray}}
\newcommand{\eeq}{\end{eqnarray}}

\newenvironment{conditions}{
\vspace{-2mm}\begin{itemize}
\setlength{\itemsep}{-1mm}
}{\vspace{-2mm}\end{itemize}}

\begin{document}

\begin{titlepage}
\begin{flushright}
 EMPG--13--11
\end{flushright}
\vskip 2.0cm
\begin{center}
{\LARGE \bf Six-Dimensional (1,0) Superconformal Models\\[0.2cm] and Higher Gauge Theory}
\vskip 1.5cm
{\Large Sam Palmer and Christian S\"amann}
\setcounter{footnote}{0}
\renewcommand{\thefootnote}{\arabic{thefootnote}}
\vskip 1cm
{\em Maxwell Institute for Mathematical Sciences\\
Department of Mathematics, Heriot-Watt University\\
Colin Maclaurin Building, Riccarton, Edinburgh EH14 4AS, U.K.}\\[0.5cm]
{Email: {\ttfamily sap2@hw.ac.uk~,~c.saemann@hw.ac.uk}}
\end{center}
\vskip 1.0cm
\begin{center}
{\bf Abstract}
\end{center}
\begin{quote}
We analyze the gauge structure of a recently proposed superconformal field theory in six dimensions. We find that this structure amounts to a weak Courant-Dorfman algebra, which, in turn, can be interpreted as a strong homotopy Lie algebra. This suggests that the superconformal field theory is closely related to higher gauge theory, describing the parallel transport of extended objects. Indeed we find that, under certain restrictions, the field content and gauge transformations reduce to those of higher gauge theory. We also present a number of interesting examples of admissible gauge structures such as the structure Lie 2-algebra of an abelian gerbe, differential crossed modules, the 3-algebras of M2-brane models and string Lie 2-algebras.
\end{quote}
\end{titlepage}

\section{Introduction and results}

The M5-branes arising in M-theory motivate the search for six-dimensional (2,0) superconformal non-abelian field theories, which have been shown to exist in \cite{Witten:1995zh}. Considerable progress has been made recently towards the construction of such a theory, following many avenues of approach. In this paper, we relate the gauge structure appearing in an approach based on tensor hierarchies in supergravity \cite{Samtleben:2011fj} to various algebraic structures appearing in the context of categorification, such as Courant algebroids, Courant-Dorfman algebras, differential crossed modules, differential 2-crossed modules, strong homotopy Lie algebras and string Lie 2-algebras.

The six-dimensional model of \cite{Samtleben:2011fj} exhibits $\CN=(1,0)$ superconformal invariance, and its field content comprises, besides the usual gauge potential one-form $A$, also gauge potential 2- and 3-forms $B$ and $C$, all taking values in a priori different vector spaces. A non-abelian action of $A$ onto $B$ and $C$ is defined, together with various other algebraic structures on the three vector spaces. The analysis of \cite{Samtleben:2011fj} led to a list of constraints on these algebraic structures necessary for closure of the (1,0) supersymmetry algebra and, in some cases, for an action to be formulated, see also \cite{Samtleben:2012mi,Samtleben:2012fb,Bandos:2013jva}. These constraints can be regarded as generalizations of the familiar Jacobi identity of Lie algebras. A special case of these theories contains the $\sG\times \sG$-model proposed in \cite{Chu:2011fd}, to which an action and interesting solutions have been constructed in \cite{Chu:2012um,Chu:2013hja,Chu:2013joa}. For solutions, such as solitons, in the general (1,0) model, see \cite{Akyol:2012cq}.

To allow for an interpretation of this (1,0) model in the context of M5-branes, it is necessary that it describes the parallel transport of extended objects. The latter is known to be nicely captured by higher gauge theory, see e.g.\ \cite{Baez:2002jn,Baez:2010ya}, and we therefore wish to relate the (1,0) model to higher gauge theory. 

A first step in this direction is an analysis of the (1,0) gauge structure. We start by noting that it forms a differential graded Leibniz algebra. Restricting the (1,0) gauge structure to an interesting class of examples, we find exact agreement of the resulting structure with Courant-Dorfman algebras \cite{Roytenberg:0902.4862}. Moreover, a general (1,0) gauge structure is a weak Courant-Dorfman algebra in the sense of \cite{Ekstrand:2009qz}. We investigate the possibility that these arise from Voronov's derived bracket construction \cite{Voronov:math0304038}, unfortunately this does not seem to be the case.

Weak Courant-Dorfman algebras, and in particular (1,0) gauge structures have a large overlap with strong homotopy Lie algebras or semistrict Lie $n$-algebras that replace gauge algebras in the context of higher gauge theory. We find that (1,0) gauge structures corresponding to Courant-Dorfman algebras form Lie 2-algebras, while many another interesting classes form Lie 3-algebras or can be extended to Lie 4-algebras. This establishes, at least in part, the desired relation to higher gauge theory. 

To strengthen the link between the (1,0) model and higher gauge theory further, we continue by studying a number of examples. The connective structure of an abelian gerbe, which underlies abelian higher gauge theory, is easily identified as a special case of the gauge potentials of the (1,0) model. Similarly, we discover the gauge algebraic structures as well as the field content and the gauge transformations of special classes of principal 2- and principal 3-bundles in the (1,0) model, establishing an overlap of the (1,0) model with strict higher gauge theory. We thus have to conclude that (1,0) models do not allow for general differential crossed and 2-crossed modules as higher gauge algebras.

Interestingly, we even recover the 3-algebras of M2-brane models as a special class of (1,0) gauge structures. We pointed out such a link between M5- and M2-brane models already in the context of crossed modules \cite{Palmer:2012ya}, which is expected because at least the BPS subsectors of the two kinds of models seem to be linked by Nahm-type transform, cf.\ \cite{Saemann:2010cp,Palmer:2011vx}.

We briefly comment on a number of further examples. First, we show how to recover both the gauge algebra as well as the action of gauge transformations of the $\sG\times \sG$-model proposed in \cite{Chu:2011fd} from the (1,0) model. Then we show that two canonical examples in higher gauge theory, the string Lie algebra of a simple Lie algebra and the Chern-Simons Lie 3-algebra of $\au(1)$ both form (1,0) gauge structures. Finally, we consider the two extreme examples of Courant-Dorfman algebras.

An interesting open question remaining is the comparison of the equations of motion of the (1,0) model to the superconformal (2,0) equations that can be obtained from a twistor construction, cf.\ \cite{Saemann:2011nb,Saemann:2012uq,Mason:2012va,Saemann:2013pca}. However, the fact that the (1,0) model makes use of structures that are only accessible in the semistrict case suggests that the twistor constructions should first be extended to principal 2-bundles with semistrict gauge 2-algebras.

This paper is structured as follows. In section 2, we review the (1,0) model and its underlying gauge structure to the extent necessary for our analysis. In section 3, we show that (1,0) gauge structures form differential graded Leibniz algebras, which are given by weak Courant-Dorfman algebras. It is then shown in section 4 that these weak Courant-Dorfman algebras form certain strong homotopy Lie algebras. Examples of (1,0) gauge structures that are closely related to higher gauge theory are then discussed in detail in section 5. An appendix recalls some basics on strong homotopy Lie algebras.

\

\noindent {\bf Remark.} While finalizing the draft of this paper, we became aware of the upcoming work of Sylvain Lavau, Henning Samtleben and Thomas Strobl \cite{Lavau:2013aa} on closely related topics.

\section{The (1,0) model}

In this section, we will briefly review the recently derived superconformal field theories in six dimensions with $\CN=(1,0)$ supersymmetry \cite{Samtleben:2011fj}. We will focus on the gauge structure, but also list the field content, gauge transformations as well as the equations of motion. 

\subsection{(1,0) gauge structures}

Consider two vector spaces $\frg$ and $\frh$ together with two linear maps $\sfg:\frg^*\rightarrow \frh$ and $\sfh: \frh\rightarrow \frg$, where $\frg^*$ denotes the dual of $\frg$. Demanding that $\sfh\circ \sfg=0$, we obtain the chain complex
\begin{equation}\label{eq:chain_complex}
 \frg^*\ \stackrel{\sfg}{\longrightarrow}\ \frh\ \stackrel{\sfh}{\longrightarrow}\ \frg~.
\end{equation}
We will denote elements of $\frg^*$, $\frh$ and $\frg$ by $\lambda$, $\chi$ and $\gamma$, respectively. Assume that we have further bilinear maps 
\begin{equation}\label{eq:maps_on_g_h}
 \sff:\frg\wedge \frg \rightarrow \frg~,~~~\sfd: \frg\odot \frg\rightarrow \frh~,~~~\sfb:\frh\otimes \frg\rightarrow \frg^*~.
\end{equation}
We also have the dual maps 
\begin{equation}
 \sfg^*: \frh^*\rightarrow \frg~,~~~\sfh^*: \frg^*\rightarrow \frh^*~,
\end{equation}
and, by considering one of the arguments as a parameter,
\begin{equation}
  \sff^*: \frg\times\frg^*\rightarrow \frg^*~,~~~\sfd^*:\frh^*\times \frg\rightarrow \frg^*~.
\end{equation}

We demand that all these maps satisfy the following equations\footnote{Here and in the following, we use $(\cdots)$ and $[\cdots]$ to denote weighted symmetrization and antisymmetrization of the enclose indices.} \cite{Samtleben:2011fj}:
\begin{subequations}\label{eq:algebra_relations}
\begin{equation}
2(\sfd(\sfh(\sfd(\gamma_1,\gamma_{(2})),\gamma_{3)})-\sfd(\sfh(\sfd(\gamma_2,\gamma_3)),\gamma_1))=2\sfd(\sff(\gamma_1,\gamma_{(2}),\gamma_{3)})-\sfg(\sfb(\sfd(\gamma_2,\gamma_3),\gamma_1))~,\label{eq:algebra_relation_a}
\end{equation}
\begin{equation}
 \begin{aligned}
    \sfd^*(\sfh^*(\sfb(\chi,\gamma_2)),\gamma_1)\,+\,&\sfb(\chi,\sfh(\sfd(\gamma_1,\gamma_2)))+2\sfb(\sfd(\gamma_1,\sfh(\chi)),\gamma_2)=\\&\sff^*(\gamma_1,\sfb(\chi,\gamma_2))+\sfb(\chi,\sff(\gamma_1,\gamma_2))+\sfb(\sfg(\sfb(\chi,\gamma_1)),\gamma_2)\label{eq:algebra_relation_b}
 \end{aligned}
\end{equation}
and
\begin{eqnarray}
 \sfh(\sfg(\lambda))&=&0~,\label{eq:algebra_relation_c}\\
 \sff(\sfh(\chi),\gamma)-\sfh(\sfd(\sfh(\chi),\gamma))&=&0~,\label{eq:algebra_relation_d}\\
 \sff(\gamma_{[1},\sff(\gamma_2,\gamma_{3]}))-\tfrac{1}{3}\sfh(\sfd(\sff(\gamma_{[1},\gamma_2),\gamma_{3]}))&=&0~,\label{eq:algebra_relation_e}\\
 \sfg(\sfb(\chi_1,\sfh(\chi_2)))-2\sfd(\sfh(\chi_1),\sfh(\chi_2))&=&0~,\label{eq:algebra_relation_f}\\
 \sfg(\sff^*(\gamma,\lambda)-\sfd^*(\sfh^*(\lambda),\gamma)+\sfb(\sfg(\lambda),\gamma))&=&0~.\label{eq:algebra_relation_g}
\end{eqnarray}
\end{subequations}
We will refer to such a structure, i.e.\ a chain complex \eqref{eq:chain_complex} together with maps \eqref{eq:maps_on_g_h} satisfying \eqref{eq:algebra_relations} as a {\em (1,0) gauge structure}.

As an initial remark, note that the map $ \sff:\frg\wedge \frg \rightarrow \frg$ is very similar to a Lie bracket on $\frg$, with \eqref{eq:algebra_relation_e} showing the failure of the Jacobi identity to hold.

Equations \eqref{eq:algebra_relations} guarantee that there is a Lie algebra $\CA$ isomorphic to $\frg$ as a vector space that has the following two representations on $\frg$ and $\frh$:
\begin{subequations}\label{eq:reps_A}
\begin{equation}
 \rho(X)\acton \gamma=-\sff(X,\gamma)+\sfh(\sfd(X,\gamma))~,
\end{equation}
and 
\begin{equation}
 \rho(X)\acton \chi:=2\sfd(X,\sfh(\chi))-\sfg(\sfb(\chi,X))
\end{equation}
for $X\in\CA$. The representation on $\frg$ also induces a representation on $\frg^*$,
\begin{equation}
 \rho(X)\acton \lambda=\sff^*(X,\lambda)-\sfd^*(\sfh^*(\lambda),X)~.
\end{equation}
\end{subequations}
All the representations satisfy the relation\footnote{Note that equations \eqref{eq:reps_A} and \eqref{eq:rep_commutator} define the Lie algebra $\CA$ only up to representations. Unless one of them is faithful, there is no unique Lie algebra structure on $\CA$ that could be reconstructed.}
\begin{equation}\label{eq:rep_commutator}
 [\rho(X_1),\rho(X_2)]=\rho(-\sff(X_1,X_2)+\sfh(\sfd(X_1,X_2)))=\rho(-\sff(X_1,X_2))~.
\end{equation}
Finally, all the maps introduced above are invariant under the action of $\CA$ because equations
\begin{subequations}
\begin{eqnarray}
 \rho(X)\acton \sfd(\gamma_1,\gamma_2)&=&\sfd(\rho(X)\acton\gamma_1,\gamma_2)+\sfd(\gamma_1,\rho(X)\acton\gamma_2)~,\\
 \rho(X)\acton \sfb(\chi,\gamma)&=&\sfb(\rho(X)\acton\chi,\gamma)+\sfb(\chi,\rho(X)\acton\gamma)~,\\
 \rho(X)\acton \sfh(\chi)&=&\sfh(\rho(X)\acton\chi)~,\label{eq:rep_rel_c}\\
 \rho(X)\acton \sfg(\lambda)&=&\sfg(\rho(X)\acton\lambda)\label{eq:rep_rel_d}
\end{eqnarray}
\end{subequations}
are equivalent to \eqref{eq:algebra_relation_a}, \eqref{eq:algebra_relation_b}, \eqref{eq:algebra_relation_d} and \eqref{eq:algebra_relation_g}, respectively. Furthermore, the invariance of $\sff$ implies \eqref{eq:algebra_relation_e}.

To analyze the above equations further, one can choose a convenient basis for $\frg$ and $\frh$, in which either the map $\sfg$ or $\sfh$ is diagonal as was done in \cite{Samtleben:2012mi}.

If one demands that the (1,0) model allows for an action principle, one has to require in addition that there is a nondegenerate bilinear form $(\cdot,\cdot)_\frh$ on $\frh$, which induces a linear nondegenerate map $\sfm:\frh\rightarrow \frh^*$ with $\sfm\circ \sfm^*=\sfm^*\circ\sfm=\mathsf{id}$. Furthermore, the following conditions have to be satisfied:
\begin{subequations}\label{eq:algebra_relations_action}
\begin{eqnarray}
  \sfg(\lambda)&=&\sfm^*(\sfh^*(\lambda))~,\label{eq:algebra_relation_action_a}\\
  \sfb(\chi,\gamma)&=&2\sfd^*(\sfm(\chi),\gamma)~,\label{eq:algebra_relation_action_b}\\
  \big(\sfd(\gamma_1,\gamma_{(2}),\sfd(\gamma_2,\gamma_{3)})\big)_\frh&=&0~.\label{eq:algebra_relation_action_c}
\end{eqnarray}
\end{subequations}
Below, we will impose the additional relations \eqref{eq:algebra_relations_action} only if explicitly stated.

\subsection{Field content}

The field content of the superconformal (1,0) theory is given by a gauge potential one-form $A$ taking values in $\frg$, a two-form potential $B$ taking values in $\frh$ and a three-form potential $C$ with values in $\frg^*$. Their curvatures read as 
\begin{subequations}\label{eq:curvatures}
\begin{eqnarray}
\CF&=&\dpar  A-\tfrac{1}{2}\sff(A,A)+\sfh(B)~,\label{eq:curvatureF10}\\
\CH&=&DB+\sfd(A,\dpar  A-\tfrac{1}{3}\sff(A,A))+\sfg(C)\nonumber\\
&=&\dpar  B+ 2\sfd(A,\sfh(B))-\sfg(\sfb(B,A))+\sfd(A,\dpar  A-\tfrac{1}{3}\sff(A,A))+\sfg(C)~\label{eq:curvatureH10},
\end{eqnarray}
\end{subequations}
where, to avoid confusion with the map $\sfd:\frg\odot \frg\rightarrow \frh$ , we will use $\dpar$ for the exterior derivative throughout this paper, e.g.
\begin{equation}
\dpar  A:=\dpar_{[\mu}A_{\nu]}\dd x^\mu\wedge\dd x^\nu~.
\end{equation} 
The covariant derivative acts by $D=\dpar +\rho(A)\acton$ and, in our notation, maps acting on the (1,0) gauge structure do not act on the form part of the fields, e.g.
\begin{equation}
 \sff(A,A):=\sff(A_\mu,A_\nu)\dd x^\mu\wedge\dd x^\nu~.
\end{equation}

Infinitesimal gauge transformations are parametrized by a function $\alpha$ taking values in $\frg$, as well as 1- and 2-forms $\Lambda$ and $\Xi$ with values in $\frh$ and $\frg^*$, respectively. Their action on the potential forms are
\begin{equation}\label{eq:qauge}
\begin{aligned}
\delta A&=D\alpha-\sfh(\Lambda)~,\\
\delta B&=D \Lambda+\sfd(A,D \alpha -\sfh(\Lambda))-2\sfd(\alpha,\CF)-\sfg(\Xi)~,\\
\delta  C&=D \Xi+\sfb(B,D\alpha-\sfh(\Lambda))-\tfrac{1}{3}\sfb(\sfd(D\alpha-\sfh(\Lambda),A),A)+\sfb(\Lambda,\CF)+\sfb(\CH,\alpha)+\dots~,
\end{aligned}
\end{equation}
where $\dots$ represents further terms in the kernel of $\sfg$. Later, we will find it useful to use a shifted version of these gauge transformations. Taking the shifted parameters $(\alpha,\Lambda,\Xi)\rightarrow(\alpha,\Lambda+\sfd(\alpha,A), \Xi-\sfb(B,\alpha)+\tfrac{1}{3}\sfb(\sfd(\alpha,A),A))$ we obtain
\begin{equation}\label{eq:shiftqauge}
\begin{aligned}
\delta A=&~\dpar  \alpha-\sff(A,\alpha)-\sfh(\Lambda)~,\\
\delta B=&~\dpar  \Lambda+\sfd(A,\sfh(\Lambda))+\sfg(\sfb(\Lambda,A))-\sfd(\alpha,\sfh(B))+\sfg(\sfb(B,\alpha))-\sfg(\Xi)\\
&-\sfd(\alpha,\CF)+\tfrac{1}{6}(\sfd(\sff(A,A),\alpha)+2\sfd(\sff(A,\alpha),A))~,\\
\delta  C=&~\dpar  \Xi-\sfb(\dpar  B,\alpha)+\tfrac{1}{3}(\sfb(\sfd(\alpha,\dpar  A),A)-\sfb(\sfd(\alpha,A),\dpar  A))\\&-\sfb(\sfg(\Xi-\sfb(B,\alpha)+\tfrac{1}{3}\sfb(\sfd(\alpha,A),A))),A)\\&+\sfb(B,-\sff(A,\alpha)-\sfh(\Lambda))-\tfrac{1}{3}\sfb(\sfd(-\sff(A,\alpha)-\sfh(\Lambda),A),A)\\&+\sfb(\Lambda+\sfd(\alpha,A),\CF)+\sfb(\CH,\alpha)+\dots~,
\end{aligned}
\end{equation}
where we used \eqref{eq:algebra_relation_g} and \eqref{eq:algebra_relation_a} in the form of
\begin{equation}\label{eq:forma}
\begin{aligned}
\sfd(A,\sff(A,\alpha)-3\sfh(\sfd(A,\alpha))-\sfd(\alpha,\sff(A,A))=\sfg(\sfb(\sfd(\alpha,A),A))~.
\end{aligned}
\end{equation}

\subsection{Bianchi identities and extended complexes}\label{ssec:Bianchi_Extension}

By construction, the field strengths satisfy the Bianchi identity
\begin{equation}
\begin{aligned}
D \CF=\sfh(\CH)~.
\end{aligned}
\end{equation}
Furthermore, demanding that 
\begin{equation}
\begin{aligned}
D \CH=\sfd(\CF,\CF)+\sfg(\CH^{(4)})~,
\end{aligned}
\end{equation}
for some four-form $\CH^{(4)}$, defined up to terms in the kernel of $\sfg$, leads to 
\begin{equation}
\begin{aligned}
D \CH^{(4)}=\sfb(\CH,\CF)+\dots~,
\end{aligned}
\end{equation}
where $\dots$ again represents terms in the kernel of $\sfg$. 

This process can be continued by extending the complex\footnote{Such an extension can always be found; for example, we could put $\frl=\ker(\sfg)$ and $\sfk$ is its embedding into $\frg$.}
\begin{equation}
\frl\overset{\sfk}{\longrightarrow} \frg^*\longrightarrow \frh\longrightarrow \frg ~,
\end{equation}
and defining a five-form $\CH^{(5)}\in\frl$ such that 
\begin{equation}
\begin{aligned}
D \CH^{(4)}=\sfb(\CH,\CF)+\sfk(\CH^{(5)})~,
\end{aligned}
\end{equation}
and such that $\CH^{(5)}$ satisfies its own Bianchi identity involving new maps into $\frl$ which satisfy additional constraints. These are found in \cite{Samtleben:2011fj} and \cite{Bandos:2013jva}. In the latter paper this extended model was used to write down a PST-like action. This extension is very similar to that of higher gauge theory with iterated categorifications of principal bundles. In the following, however, we will restrict ourselves to the non-extended case.

\subsection{Supersymmetry and field equations}

For this section we will introduce the notation 
\begin{equation}
\begin{aligned}
\gamma&=\gamma_\mu\dd x^\mu~,~~&\gamma^{(2)}&=\tfrac{1}{2}\gamma_{\mu\nu}\dd x^\mu\wedge\dd x^\nu~,~~&\gamma^{(3)}&=\tfrac{1}{6}\gamma_{\mu\nu\rho}\dd x^\mu\wedge\dd x^\nu\wedge\dd x^\rho~,\\
\slasha{D}&=\gamma^\mu D_\mu~,~~&\slasha{\CF}&=*(\CF\wedge*\gamma^{(2)})=\tfrac{1}{2}\gamma^{\mu\nu}\CF_{\mu\nu}~,~&\slasha{\CH}&=*(\CH\wedge*\gamma^{(3)})=\tfrac{1}{6}\gamma^{\mu\nu\rho}\CH_{\mu\nu\rho}~,
\end{aligned}
\end{equation}
where $*$ is the Hodge star operation. The fields above belong to the (1,0) vector and tensor supermultiplets $(A,\lambda^i,Y^{ij})$ and $(\phi,\chi^i,B)$, for $i,j=1,2$, taking values in $\frg$ and $\frh$, respectively. In \cite{Samtleben:2011fj}, it was found that the supersymmetry transformations 
\begin{equation}\label{eq:susy}
\begin{aligned}
\delta A&=-\epsb \gamma \lambda~,~~~&\delta B&=-\sfd(A,\epsb \gamma \lambda)-\epsb\gamma^{(2)}\chi~,\\
\delta \lambda^i&=\tfrac{1}{4}\slasha{\CF}\eps^i-\tfrac{1}{2}Y^{ij}\eps_j+\tfrac{1}{4}\sfh(\phi)\eps^i~,~~~&\delta \chi^i&=\tfrac{1}{8}\slasha{\CH}\eps^i+\tfrac{1}{4}\slasha{D}\phi~\eps^i-*\tfrac{1}{2}\sfd(\gamma\lambda^i,*\epsb\gamma\lambda)~,\\
\delta  Y^{ij}&=-\epsb^{(i}\slasha{D} \lambda^{j)}+2 \epsb^{(i}\sfh(\chi^{j)})~,~~~&\delta \phi&=\epsb\chi~,\\
&&&\hspace{-4.3cm}\delta C=-\sfb(B,\epsb \gamma\lambda)-\tfrac{1}{3}\sfb(\sfd(A,\epsb \gamma \lambda), A)-\sfb(\phi,\epsb\gamma^{(3)}\lambda)~,
\end{aligned}
\end{equation}
close up to translations, gauge transformations and the equations of motion
\begin{equation}\label{eq:eom}
\begin{aligned}
\CH^-&=-\sfd(\lambdab,\gamma^{(3)}\lambda)~,\\
\slasha{D} \chi^i&=\sfd(\slasha{\CF},\lambda^i)+2\sfd(Y^{ij},\lambda_j)+\sfd(\sfh(\phi),\lambda^i)-2\sfg(\sfb(\phi,\lambda^i))~,\\
D^2 \phi&=2\sfd(Y^{ij},Y_{ij})-*2\sfd(\CF,*\CF)-4\sfd(\lambdab,\slasha{D} \lambda)\\&~~~~-2\sfg(\sfb(\chib,\lambda))+16\sfd(\lambdab,\sfh(\chi))-3\sfd(\sfh(\phi),\sfh(\phi))~,
\end{aligned}
\end{equation}
where $\CH=\CH^++\CH^-$ is split into selfdual and anti-selfdual parts: $\CH^\pm=\pm*\CH^\pm$. These tensor multiplet equations \eqref{eq:eom} are connected by supersymmetry to the following vector multiplet equations 
\begin{equation}\label{eq:eom2}
\begin{aligned}
\sfg(\sfb(\phi,Y_{ij})+2\sfb(\chib_{(i},\lambda_{j)}))&=0~,\\
\sfg(\sfb(\phi,\CF)-2\sfb(\chib,\gamma^{(2)}\lambda))&=\tfrac{1}{2}\sfg(*\CH^{(4)})~,\\
\sfg(\sfb(\phi,\slasha{D} \lambda_i)+\tfrac{1}{2}\sfb(\slasha{D}\phi, \lambda_i))&=\sfg(*\tfrac{1}{2}\sfb(\gamma^{(2)}\chi_i,*\CF)+\tfrac{1}{4}\sfb(\slasha{\CH},\lambda_i)-\sfb(\chi^j,Y_{ij})\\&~~~~+\tfrac{3}{2}\sfb(\phi,\sfh(\chi))+*\tfrac{1}{3}\sfb(\sfd(\gamma\lambda_i,\lambdab),*\gamma\lambda))~.
\end{aligned}
\end{equation}

\section{(1,0) gauge structures and weak Courant-Dorfman algebras}

\subsection{Differential graded Leibniz algebra}

We now come to the analysis of the gauge structure that is defined by the maps \eqref{eq:maps_on_g_h} together with equations \eqref{eq:algebra_relations}. The fact that underlying the (1,0) gauge structure is the chain complex \eqref{eq:chain_complex} suggests that we are working with some differential graded algebra\footnote{For a detailed analysis of the general tensor hierarchy algebra from the perspective of Lie superalgebras, see \cite{Palmkvist:2013vya}.}. We first focus on the representations of the Lie algebra $\CA$ \eqref{eq:reps_A} on the vector spaces $\frg$, $\frh$ and $\frg^*$. As they satisfy the Jacobi identity, we arrive at a Leibniz algebra.

Recall that a {\em differential graded Leibniz algebra}\footnote{or a {\em differential graded Loday algebra}} $(L,\CD,\acton)$ is a ($\RZ$-)graded vector space $L$ equipped with a degree $1$ linear map $\CD$ and a degree $0$ bilinear map $\acton$ such that 
\begin{conditions}
 \item[(i)] $\CD$ is a differential: $\CD^2=0$ and $\CD(\ell_1\acton \ell_2)=(\CD \ell_1)\acton \ell_2+(-1)^{|\ell_1|}\ell_1\acton (\CD \ell_2)$~,
 \item[(ii)] a Leibniz identity holds: $\ell_1\acton(\ell_2\acton \ell_3)=(\ell_1\acton \ell_2)\acton \ell_3+(-1)^{|\ell_1||\ell_2|}\ell_2\acton(\ell_1\acton \ell_3)$~,
\end{conditions}
where $\ell_1,\ell_2,\ell_3\in L$ and $|\ell_i|$ denotes the grading of $\ell_i$. 

In the case of a (1,0) gauge structure, we have\footnote{Recall that $V[-n]$ denotes the vector space $V$ shifted by $-n$ degrees in the grading. In particular, $\frg^*[-2]$ consists of elements in $\frg^*$, and each element has homogeneous grading -2.}
\begin{equation}
 L=\frg^*[-2]\oplus \frh[-1]\oplus \frg~,~~~\CD|_{\frg^*}=\sfg~,~~~\CD|_{\frh}=\sfh~,
\end{equation}
and the only nontrivial actions $\acton$ are given by \eqref{eq:reps_A}:
\begin{equation}
 \gamma_1\acton \gamma_2:=\rho(\gamma_1)\acton \gamma_2~,~~~\gamma_1\acton \chi:=\rho(\gamma_1)\acton \chi~,~~~\gamma_1\acton \lambda:=\rho(\gamma_1)\acton \lambda
\end{equation}
for all $\gamma_1,\gamma_2\in\frg$, $\chi\in\frh$ and $\lambda\in\frg^*$. Conditions (i) and (ii) are readily verified: (i) follows from \eqref{eq:algebra_relation_c} together with \eqref{eq:rep_rel_c} and \eqref{eq:rep_rel_d}, while (ii) follows from the fact that $\rho$ forms a representation of $\CA$.

The characterization of (1,0) gauge algebras in terms of Leibniz algebras is certainly too general. In particular, we would like to identify a structure in which the maps $\sff$, $\sfd$ and $\sfb$ are given an intrinsic meaning. Clearly, considering separately the antisymmetrization and the symmetrization of 
\begin{equation}
 \gamma_1\acton\gamma_2:=\rho(\gamma_1)\acton\gamma_2=-\sff(\gamma_1,\gamma_2)+\sfh(\sfd(\gamma_1,\gamma_2))
\end{equation}
would allow us to extract $\sff$ as well as $\sfd$ up to terms in the kernel of $\sfh$. Note, however, that these new maps cannot be expected to satisfy the Leibniz identity anymore. The transition between a product satisfying a Leibniz identity and its antisymmetrization that violates the Leibniz rule (which here amounts to the Jacobi identity) is in fact a very common one in the context of Courant algebroids. We therefore turn our attention to those in the following.

\subsection{Courant algebroids}

A particularly nice class of examples of (1,0) gauge structures is obtained from Courant algebroids. Recall that a Courant algebroid is a symplectic Lie 2-algebroid, or, equivalently, a symplectic NQ-manifold\footnote{a Q-manifold with non-negatively integer grading which is endowed with a symplectic form}, cf.\ \cite{Roytenberg:0203110}. Here, we define it as a Euclidean vector bundle $(E, \langle \cdot,\cdot\rangle)$ over a smooth manifold $M$ that is endowed with a bilinear operation $\acton$ on sections of $E$ and a bundle map $\varrho:E\rightarrow TM$ called the {\em anchor} satisfying the following axioms for all $e,e_1,e_2\in \Gamma(E)$ and $f\in\CC^\infty(M)$:
\begin{conditions}
 \item[(i)] $e\acton(e_1\acton e_2)=(e\acton e_1)\acton e_2+e_1\acton(e\acton e_2)$,
 \item[(ii)] $e_1\acton e_2+e_2\acton e_1=\CD\langle e_1,e_2\rangle$,
 \item[(iii)] $\varrho(e_1\acton e_2)=[\varrho(e_1),\varrho(e_2)]$,
 \item[(iv)] $e_1\acton (fe_2)=f(e_1\acton e_2)+(\varrho(e_1)\cdot f)e_2$,
 \item[(v)] $\varrho(e)\cdot \langle e_1,e_2\rangle=\langle e\acton e_1,e_2\rangle+\langle e_1,e\acton e_2\rangle$.
\end{conditions}
Here $\varrho(e)\cdot f$ denotes the action of the vector field $\varrho(e)$ onto $f$, $[\cdot,\cdot]$ denotes the Lie bracket of vector fields and $\CD$ is the pullback of the exterior derivative $\dpar$ on $M$ via the adjoint map $\varrho^*$:
\begin{equation}
 \langle \CD f,e \rangle:=\tfrac{1}{2}\varrho(e)\cdot f~.
\end{equation}

A Courant algebroid contains a differential graded Leibniz algebra, and one can show that it forms a (1,0) gauge structure with trivial maps $\sfg$ and $\sfb$. Instead of doing this using the above definition, which stems from \cite{Roytenberg:0203110}, we can switch to the original and equivalent definition from \cite{Liu:1997aa}. For this, we introduce the antisymmetric {\em Courant bracket}
\begin{equation}\label{eq:switch_brackets}
\llbracket e_1,e_2\rrbracket:=\tfrac{1}{2}(e_1\acton e_2-e_2\acton e_1)=e_1\acton e_2-\tfrac{1}{2}\CD\langle e_1,e_2\rangle~.
\end{equation}
In this context, the action $\acton$ is often called a {\em Dorfman bracket}. For the Courant bracket, the axioms in the definition of a Courant algebroid become
\begin{conditions}
 \item[(i')] $\llbracket\llbracket e_1,e_2\rrbracket,e_3\rrbracket+\llbracket\llbracket e_2,e_3\rrbracket, e_1\rrbracket+\llbracket\llbracket e_3,e_1\rrbracket,e_2\rrbracket+\tfrac{1}{2}\CD\big\langle\llbracket e_{[1},e_2\rrbracket, e_{3]}\big\rangle=0$,
 \item[(iii')] $\varrho(\llbracket e_1,e_2\rrbracket)=[\varrho(e_1),\varrho(e_2)]$,
 \item[(iv')] $\llbracket e_1,f e_2\rrbracket=f\llbracket e_1,e_2\rrbracket+(\varrho(e_1)\cdot f)e_2-\langle e_1,e_2\rangle\CD f$,
 \item[(v')] $\varrho(e)\cdot\langle e_1,e_2\rangle =\big\langle\llbracket e,e_1\rrbracket+\CD\langle e,e_1\rangle,e_2 \big\rangle+ \big\langle e_1 ,\llbracket e,e_2\rrbracket+\CD\langle e ,e_2\rangle \big\rangle$,
 \item[(vi')] $\langle \CD f,\CD g\rangle=0$,
\end{conditions}
where again $e,e_1,e_2\in \Gamma(E)$ and $f,g\in \CC^\infty(M)$.

Given a Courant algebroid, we can define a (1,0) gauge structure by putting
\begin{equation}
 \begin{aligned}
    \frg:=\Gamma(E)~,~~~\frh:=\CC^\infty(M)~,~~~\sfh:=\CD~,~~~\sff:=-\llbracket\cdot,\cdot\rrbracket~,~~~\sfd:=\tfrac{1}{2}\langle \cdot,\cdot \rangle~,~~~ \sfg:=0~,~~~\sfb:=0~.
 \end{aligned}
\end{equation}
The relations \eqref{eq:algebra_relation_b}, \eqref{eq:algebra_relation_c}, \eqref{eq:algebra_relation_g} are trivially satisfied. Moreover, the relations \eqref{eq:algebra_relation_a}, \eqref{eq:algebra_relation_e} and \eqref{eq:algebra_relation_f} are equivalent to the axioms (v'), (i') and (vi'), respectively. Finally, equation \eqref{eq:algebra_relation_d} has been shown to hold for Courant algebroids \cite[Prop. 4.2]{Roytenberg:1998vn}.

To capture finite dimensional (1,0) gauge structures, we need to reformulate the notion of a Courant algebroid in purely algebraic terms. This leads to the concept of a Courant-Dorfman algebra.

\subsection{Courant-Dorfman algebras}\label{ssec:Courant_Dorfman_algebras}

A {\em Courant-Dorfman algebra} \cite{Roytenberg:0902.4862}, see also \cite{Keller:0807.0584}, consists of a commutative $\mathbb{K}$-algebra $\CR$ together with an $\CR$-module $\CE$ endowed with a derivation $\CD:\CR\rightarrow \CE$, a symmetric bilinear form (not necessarily non-degenerate) $\langle\cdot,\cdot\rangle: \CE\otimes_\CR\CE\rightarrow \CE$ and a {\em Dorfman bracket} $\acton:\CE\otimes \CE\rightarrow \CE$, which satisfy the following axioms:
\begin{conditions}
 \item[(i)] $e_1\acton (e_2\acton e_3)=(e_1\acton e_2)\acton e_3+e_2\acton(e_1\acton e_3)$,
 \item[(ii)] $e_1\acton e_2+e_2\acton e_1=\CD\langle e_1,e_2\rangle$,
 \item[(iii)] $(\CD r)\acton e=0$,
 \item[(iv)] $e_1\acton r e_2=r(e_1\acton e_2)+\langle e_1,\CD r\rangle e_2$,
 \item[(v)] $\langle e_1,\CD\langle e_2,e_3\rangle\rangle=\langle e_1\acton e_2,e_3\rangle+\langle e_2,e_1\acton e_3\rangle$,
 \item[(vi)] $\langle \CD r_1,\CD r_2\rangle=0$,
\end{conditions}
where $e,e_1,e_2,e_3\in\CE$ and $r,r_1,r_2\in\CR$. Note that if the bilinear form $\langle\cdot,\cdot\rangle$ is non-degenerate, axioms (iii), (iv) and (vi) are redundant. Moreover, if we consider a Euclidean vector bundle $E\rightarrow M$ with fiber metric $\langle \cdot,\cdot\rangle$, put $\CE=\Gamma(E)$ and $\CR=\CC^\infty(M)$ and define $\CD$ as the pullback of the exterior derivative on $M$, then we recover the notion of a Courant algebroid.

As before, we can reformulate these axioms by switching from the Dorfman bracket $\acton$ to the Courant bracket via relation \eqref{eq:switch_brackets}, and we are left with
\begin{conditions}
 \item[ (i')] $\llbracket e_{[ 1},\llbracket e_2,e_{3]}\rrbracket \rrbracket +\tfrac{1}{6}\CD\langle e_{[ 1},\llbracket e_2,e_{3]}\rrbracket \rangle=0$,
 \item[ (iii')] $\llbracket \CD r,e\rrbracket +\tfrac{1}{2}\CD\langle \CD r,e\rangle=0$,
 \item[ (iv')] $\llbracket e_1,r e_2\rrbracket =r\llbracket e_1,e_2\rrbracket +\langle e_1,\CD r\rangle e_2+\tfrac{1}{2}r(\CD\langle e_1,e_2\rangle)-\tfrac{1}{2}\CD\langle e_1,r e_2\rangle$,
 \item[ (v')] $\langle \CD\langle e_1,e_{(2}\rangle,e_{3)}\rangle-\langle \CD\langle e_2,e_3\rangle,e_1\rangle+2\langle\llbracket e_1,e_{(2}\rrbracket ,e_{3)}\rangle=0$,
 \item[ (vi')] $\langle \CD r_1,\CD r_2\rangle=0$.
\end{conditions}

Given a Courant-Dorfman algebra, we can construct a (1,0) gauge structure by putting 
\vspace*{-0.6cm}
\begin{equation}
 \begin{aligned}
    \frg:=\CE~,~~~\frh:=\CR~,~~~\sfh:=\CD~,~~~\sff:=-\llbracket\cdot,\cdot\rrbracket~,~~~\sfd:=\tfrac{1}{2}\langle \cdot,\cdot \rangle~,~~~ \sfg:=0~,~~~\sfb:=0~.
 \end{aligned}
\end{equation}
Axioms \eqref{eq:algebra_relation_a}, \eqref{eq:algebra_relation_d}, \eqref{eq:algebra_relation_e} and \eqref{eq:algebra_relation_f} of the (1,0) gauge structure correspond to the axioms (v'), (iii'), (i') and (vi') of the Courant-Dorfman algebra, respectively.

Inversely, a (1,0) gauge structures with $\sfg$ and $\sfb$ trivial gives rise to a Courant-Dorfman algebra, where the action of $\CR=\frh$ onto $\CE=\frg$ is given by
\begin{equation}
r e:=\CD\langle e,\CD r\rangle=\sfh(\rho(e)\acton r)~. 
\end{equation}
Axiom (iv') holds then by definition, the other axioms are related to those of the (1,0) gauge structure as before.

\subsection{Weak Courant-Dorfman algebras}

To extend this correspondence to the case of (1,0) gauge structures with non-trivial maps $\sfg$ and $\sfb$, we have to allow for some more general Courant-Dorfman algebras. In particular, we have to weaken axioms (v') and (vi'), which correspond to \eqref{eq:algebra_relation_a} and \eqref{eq:algebra_relation_f} only for trivial $\sfg$ and $\sfb$. Interestingly, this generalization has already been introduced in \cite{Ekstrand:2009qz} by dropping axioms (iv), (v) and (vi) (or, equivalently, (iv'), (v') and (vi')) of a Courant-Dorfman algebra:

A {\em weak Courant-Dorfman algebra} consists of two vector spaces $\CR$ and $\CE$ together with a symmetric bilinear form $\langle \cdot,\cdot\rangle:\CE\otimes \CE\rightarrow \CR$, a map $\CD:\CR\rightarrow \CE$ and a Dorfman bracket $\acton:\CE\otimes \CE\rightarrow \CE$. These satisfy the following axioms:
\begin{conditions}
 \item[(i'')] $e_1\acton (e_2\acton e_3)=(e_1\acton e_2)\acton e_3+e_2\acton(e_1\acton e_3)$,
 \item[(ii'')] $e_1\acton e_2+e_2\acton e_1=\CD\langle e_1,e_2\rangle$,
 \item[(iii'')] $(\CD r)\acton e=0$.
\end{conditions}

An important class of examples is given by the higher generalizations of exact Courant algebroids $TM\oplus \wedge^p T^*M$ together with the standard Courant brackets. Since these do not seem to be related to our discussion, we refrain from going into further details.

Note that the above axioms imply the following weaker form of (v) and (vi) \cite{Ekstrand:2009qz}:
\begin{equation}
 \begin{aligned}
 \CD\big(\langle e_1,\CD\langle e_2,e_3\rangle\rangle-\langle e_1\acton e_2,e_3\rangle-\langle e_2,e_1\acton e_3\rangle\big)&=0~,\\
 \CD\langle \CD e_1,\CD e_2\rangle&=0~.
 \end{aligned}
\end{equation}
These equations are precisely the generalizations necessary to accommodate a (1,0) gauge structure with non-trivial $\sfg$ and $\sfb$, as axioms \eqref{eq:algebra_relation_a} and \eqref{eq:algebra_relation_f} are modified by terms in the image of $\sfg$, which vanishes under $\CD$ due to $\sfh\circ\sfg=0$. We therefore conclude that (1,0) gauge structures are special cases of weak Courant-Dorfman algebras.

\subsection{Comments on derived brackets}

To construct weak Courant-Dorfman algebras, one is quickly led to the notion of derived brackets: Courant algebroids are symplectic NQ-manifolds \cite{Severa:2001aa,Roytenberg:0203110}, see also \cite{Kotov:2010wr}, and $\CD$, as well as the Courant bracket $\llbracket\cdot,\cdot\rrbracket$ on sections, are derived from the symplectic structure on an NQ-manifold \cite{Roytenberg:0203110} via a derived bracket construction \cite{Kosmann-Schwarzbach:0312524,Voronov:math0304038}. Unfortunately, this approach to (1,0) gauge structures seems too restrictive, at least if one uses the superextension due to \cite{Voronov:math0304038}, as we demonstrate in the following.

We start from a Lie superalgebra $L$ with Lie bracket $\{\cdot,\cdot\}$ together with a projector $P\in \sEnd L$ onto an abelian subalgebra of $L$ such that 
\begin{equation}
 P^2=P~,~~~\{P\ell_1,P\ell_2\}=0\eand P\{\ell_1,\ell_2\}=P\{P\ell_1,\ell_2\}+P\{\ell_1,P\ell_2\}~.
\end{equation}
Given an odd element $Q\in L$ (with appropriate $\RZ$-grading) such that $Q^2=\tfrac{1}{2}\{Q,Q\}=0$, we can define the brackets
\begin{equation}
 \mu_i(\ell_1,\ell_2,\ldots,\ell_i):=P\{\ldots\{\{Q,\ell_1\},\ell_2\},\ldots,\ell_i\}~,
\end{equation}
which turn $L$ into an $L_\infty$-algebra \cite[Cor.\ 1]{Voronov:math0304038}. In particular, the condition $Q^2=0$ is equivalent to the higher homotopy relations \eqref{eq:homotopyJacobi}. Note that the grading of the $L_\infty$-algebra is again that of the Lie superalgebra shifted by one.

We now wish to identify the additional structure maps $\sfd$ and $\sfb$ with (parts of) a Poisson bracket. For this, note that equation \eqref{eq:algebra_relation_a} implies
\begin{equation}
 \sfg(\sfb(\sfd(\gamma_{(1},\gamma_2),\gamma_{3)}))=0~.
\end{equation}
If we impose either the additional constraint \eqref{eq:algebra_relation_action_c} or consider the extended tensor hierarchy (cf.\ \cite{Hartong:2009vc,Samtleben:2011fj}), one has the stronger relation
\begin{equation}\label{eq:angular_graded_jacobi}
 \sfb(\sfd(\gamma_{(1},\gamma_2),\gamma_{3)})=0~.
\end{equation}
This relation is in fact the graded Jacobi identity we require, assuming a parity shift of $\frg$ by one to odd grading. We are thus led to identify
\begin{equation}
 \{\gamma_1,\gamma_2\}=\sfd(\gamma_1,\gamma_2)\eand \{\gamma,\chi\}=\sfb(\chi,\gamma)~.
\end{equation}
If we demand in addition that
\begin{equation}
 P\{\gamma_1,\gamma_2\}=\{\gamma_1,\gamma_2\}~,
\end{equation}
then relations \eqref{eq:algebra_relation_c}, \eqref{eq:algebra_relation_d} and \eqref{eq:algebra_relation_e} are automatically satisfies, as one readily verifies. Equation \eqref{eq:algebra_relation_f} leads to a constraint:
\begin{equation}
\sfd(\sfh(\chi_1),\sfh(\chi_2))=\{P\{Q,\chi_1\},P\{Q,\chi_2\}\}=0\stackrel{!}{=}\tfrac{1}{2}\sfg(\sfb(\chi_1,\sfh(\chi_2)))~.
\end{equation}
A similar constraint is derived from \eqref{eq:algebra_relation_a}. More importantly, however, we have
\begin{equation}
 \{\mu_2(\gamma_1,\gamma_2),\mu_2(\gamma_3,\gamma_4)\}=\{P\{\{Q,\gamma_1\},\gamma_2\},P\{\{Q,\gamma_3\},\gamma_4\}\}=0~.
\end{equation}
All these constraints impose severe restrictions on the maps $\sff$, $\sfd$ and $\sfb$, which renders this approach essentially uninteresting for the construction of (1,0) gauge algebras.

\section{(1,0) gauge structures as Lie 3-algebra}

Having identified (1,0) gauge structures with weak Courant-Dorfman algebras, we would now like to make contact with higher or categorified gauge theory. As a first step towards this goal, we need to identify categorified Lie algebras in the (1,0) gauge structure. For our purposes, it suffices to restrict ourselves to so-called semistrict Lie 3-algebras. These arise from categorifying twice the notion of a Lie algebra and imposing antisymmetry on the higher products. For simplicity, we will often drop the label `semistrict' in the following. 

\subsection{Semistrict Lie 3-algebras}

Semistrict Lie 3-algebras are categorically equivalent to 3-term $L_\infty$- or strong homotopy Lie algebras \cite{Baez:2003aa}, see appendix \ref{app:SH_Lie_algebras} for the general definition of $L_\infty$-algebras. A 3-term $L_\infty$-algebra\footnote{also known as an $L_3$-algebra} is a graded vector space $L=L_{-2}\oplus L_{-1}\oplus L_0$, where $L_i$ has grading $i$, together with multilinear, totally graded antisymmetric maps
\begin{equation}\label{eq:products_Lie_3}
 \begin{aligned}
  &\mu_1:L_{-2}\rightarrow L_{-1}~,~~~&&\mu_1:L_{-1}\rightarrow L_{0}~,\\
  &\mu_2:L_{0}^{\wedge 2}\rightarrow L_0~,~~~&&\mu_2:L_{0}\wedge L_{-1}\rightarrow L_{-1}~,~~~&&\mu_2:L_{0}\wedge L_{-2}\rightarrow L_{-2}~,\\
  &\mu_2:L_{-1}^{\wedge 2}\rightarrow L_{-2}~,\\
  &\mu_3:L_0^{\wedge 3}\rightarrow L_{-1}~,~~~&&\mu_3:L_{-1}\wedge L_0^{\wedge 2}\rightarrow L_{-2}~,\\
  &\mu_4:L_0^{\wedge 4}\rightarrow L_{-2}~.
 \end{aligned}
\end{equation}
These maps satisfy a number of higher Jacobi or homotopy relations, which we list in the following. The map $\mu_1$ is a differential:
\begin{subequations}\label{eq:homotopy_relations}
\begin{equation}
  \mu_1^2(\lambda): =\mu_1(\mu_1(\lambda))=0~,\label{eq:homotopy_relation_a}
\end{equation}
and it is compatible with the products $\mu_2$:
\begin{eqnarray}
\mu_1(\mu_2(\gamma,\chi))&\!\!=\!\!&-\mu_2(\mu_1(\chi),\gamma)~,\label{eq:homotopy_relation_b}\\
\mu_1(\mu_2(\gamma,\lambda))&\!\!=\!\!&-\mu_2(\mu_1(\lambda),\gamma)~,\label{eq:homotopy_relation_c}\\
\mu_1(\mu_2(\chi_1,\chi_2))&\!\!=\!\!&\mu_2(\mu_1(\chi_1),\chi_2)+\mu_2(\mu_1(\chi_2),\chi_1)~.\label{eq:homotopy_relation_d}
\end{eqnarray}
The map $\mu_2$ satisfies a Jacobi identity up to correction terms given by $\mu_3$:
\begin{eqnarray}
 \mu_1(\mu_3(\gamma_1,\gamma_2,\gamma_3))&\!\!=\!\!&-\mu_2(\mu_2(\gamma_1,\gamma_2),\gamma_3)+\mu_2(\mu_2(\gamma_1,\gamma_3),\gamma_2)-\mu_2(\mu_2(\gamma_2,\gamma_3),\gamma_1)~,~~~\label{eq:homotopy_relation_e}\\
 \mu_1(\mu_3(\chi,\gamma_1,\gamma_2))&\!\!=\!\!&-\mu_3(\mu_1(\chi),\gamma_1,\gamma_2)-\mu_2(\mu_2(\gamma_1,\gamma_2),\chi)\nonumber\\
 &&-\mu_2(\mu_2(\chi,\gamma_1),\gamma_2)+\mu_2(\mu_2(\chi,\gamma_2),\gamma_1)~,\label{eq:homotopy_relation_f}\\
0&\!\!=\!\!&-\mu_3(\mu_1(\lambda),\gamma_1,\gamma_2)-\mu_2(\mu_2(\gamma_1,\gamma_2),\lambda)\nonumber\\
&&-\mu_2(\mu_2(\lambda,\gamma_1),\gamma_2)+\mu_2(\mu_2(\lambda,\gamma_2),\gamma_1) ~,\label{eq:homotopy_relation_g}\\
0&\!\!=\!\!&-\mu_3(\mu_1(\chi_1),\chi_2,\gamma)-\mu_3(\mu_1(\chi_2),\chi_1,\gamma)-\mu_2(\mu_2(\chi_1,\chi_2),\gamma)\nonumber\\
&&+\mu_2(\mu_2(\chi_1,\gamma),\chi_2)+\mu_2(\mu_2(\chi_2,\gamma),\chi_1)~.\label{eq:homotopy_relation_h}
\end{eqnarray}
The map $\mu_3$ is compatible with the map $\mu_2$ in the obvious way up to correction terms given by $\mu_4$:
\begin{equation}\label{eq:homotopy_relation_i}
 \begin{aligned}
 \mu_1(\mu_4(\gamma_1,&\gamma_2,\gamma_3,\gamma_4))+\mu_2(\mu_3(\gamma_1,\gamma_2,\gamma_3),\gamma_4)-\mu_2(\mu_3(\gamma_1,\gamma_2,\gamma_4),\gamma_3)\\
 &+\mu_2(\mu_3(\gamma_1,\gamma_3,\gamma_4),\gamma_2)-\mu_2(\mu_3(\gamma_2,\gamma_3,\gamma_4),\gamma_1)=\\
 &\mu_3(\mu_2(\gamma_1,\gamma_2),\gamma_3,\gamma_4))+\mu_3(\mu_2(\gamma_2,\gamma_3),\gamma_1,\gamma_4))+\mu_3(\mu_2(\gamma_3,\gamma_4),\gamma_1,\gamma_2))\\
 &+\mu_3(\mu_2(\gamma_1,\gamma_4),\gamma_2,\gamma_3))
 -\mu_3(\mu_2(\gamma_1,\gamma_3),\gamma_2,\gamma_4))-\mu_3(\mu_2(\gamma_2,\gamma_4),\gamma_1,\gamma_3))~.\\
  \end{aligned}
\end{equation}
\begin{equation}\label{eq:homotopy_relation_j}
 \begin{aligned}
\mu_2(\mu_3(\gamma_1,&\gamma_2,\gamma_3),\chi)-\mu_2(\mu_3(\chi,\gamma_1,\gamma_2),\gamma_3)+\mu_2(\mu_3(\chi,\gamma_1,\gamma_3),\gamma_2)\\
  &-\mu_2(\mu_3(\chi,\gamma_2,\gamma_3),\gamma_1)-\mu_4(\mu_1(\chi),\gamma_1,\gamma_2,\gamma_3)=\\
 &-\mu_3(\mu_2(\gamma_1,\gamma_2),\chi,\gamma_3))-\mu_3(\mu_2(\gamma_2,\gamma_3),\chi,\gamma_1))-\mu_3(\mu_2(\chi,\gamma_3),\gamma_1,\gamma_2))\\
 &-\mu_3(\mu_2(\chi,\gamma_1),\gamma_2,\gamma_3))
 +\mu_3(\mu_2(\gamma_1,\gamma_3),\chi,\gamma_2))+\mu_3(\mu_2(\chi,\gamma_2),\gamma_1,\gamma_3))~.
 \end{aligned}
\end{equation}
Finally, the map $\mu_4$ satisfies the following compatibility relation\footnote{Note that the total antisymmetrization is here equivalent to merely considering unshuffles in definition \eqref{eq:homotopyJacobi}.}
\begin{equation}\label{eq:homotopy_relation_k}
 \begin{aligned}
 \tfrac{1}{2}\mu_2(\mu_4(\gamma_{[1},\gamma_2,\gamma_3,\gamma_4),\gamma_{5]})+\mu_3(\mu_3(\gamma_{[1},&\gamma_2,\gamma_3),\gamma_4,\gamma_{5]})+\mu_4(\mu_2(\gamma_{[1},\gamma_2),\gamma_3,\gamma_4,\gamma_{5]})=0~.
 \end{aligned}
\end{equation}
\end{subequations}

A simple example of a Lie 3-algebra is that of the {\em Chern-Simons Lie 3-algebra} $\frcs_k(\frg)$ of a simple Lie algebra $\frg$, where $k\in\FR$ denotes the {\em level}. The graded vector space is $L=\FR[-2]\oplus(\FR\oplus \frg)[-1]\oplus \frg$, and we will denote elements of these spaces by $\lambda,\binom{\lambda}{\gamma}$ and $\gamma$, respectively. The non-vanishing higher products are defined as
\begin{equation}
 \begin{aligned}
  \mu_1(\lambda):=\binom{\lambda}{0}&~,~~~\mu_1\binom{\lambda}{\gamma}:=\gamma~,\\
  \mu_2(\gamma_1,\gamma_2):=[\gamma_1,\gamma_2]~,~\mu_2\left(\gamma_1,\binom{\lambda}{\gamma_2}\right)&:=\binom{k\langle\gamma_1,\gamma_2\rangle}{[\gamma_1,\gamma_2]},~
  \mu_2\left(\binom{\lambda_1}{\gamma_1},\binom{\lambda_2}{\gamma_2}\right):=2k\langle\gamma_1,\gamma_2\rangle~,\\
  \mu_3(\gamma_1,\gamma_2,\gamma_3)&:=k\langle\gamma_1,[\gamma_2,\gamma_3]\rangle~,
 \end{aligned}
\end{equation}
where $\langle\cdot,\cdot\rangle$ denotes the Killing form on $\frg$. In the following, we will discuss some special Lie 3-algebras that will later serve as examples for the gauge structure of the (1,0)-model.

\subsection{Semistrict Lie 2-algebras and string Lie 2-algebras}\label{ssec:Lie_2_algebras}

General semistrict Lie 2-algebras are obtained by considering Lie 3-algebras with trivial $L_{-2}$. This reduces the non-trivial products \eqref{eq:products_Lie_3} to the following ones:
\begin{equation}\label{eq:products_Lie_2}
 \begin{aligned}
  \mu_1:L_{-2}\rightarrow L_{-1}~,~~~\mu_1:L_{-1}\rightarrow L_{0}~,\hspace{2cm}\\
  \mu_2:L_{0}^{\wedge 2}\rightarrow L_0~,~~~\mu_2:L_{0}\wedge L_{-1}\rightarrow L_{-1}~,~~~
  \mu_3:L_0^{\wedge 3}\rightarrow L_{-1}~,
 \end{aligned}
\end{equation}
while the higher Jacobi relations reduce in an obvious manner. 

Let us specialize a little further. A semistrict Lie 2-algebra is called {\em skeletal}, if isomorphic objects are equivalent. This amounts to setting $\mu_1=0$. A nice class of skeletal semistrict Lie 2-algebras is obtained from a Lie algebra $\frg$, a vector space $V$ carrying a representation $\rho$ of $V$ and a Lie algebra cocycle with values in $V$, $c=H^3(\frg,V)$ \cite{Baez:2003aa}. As products on the 2-term complex $V\rightarrow \frg$, we define $\mu_1:=0$, $\mu_2:\frg \times \frg\rightarrow \frg$ as the Lie bracket, $\mu_2:\frg\times V\rightarrow V$ as the action of $\frg$ onto $V$ in the representation $\rho$ and $\mu_3:\frg\times\frg\times\frg\rightarrow V$ is given by the Lie algebra cocycle $c$. 

It is shown in \cite{Baez:2003aa} that isomorphism classes of such data $(\frg,V,\rho,c)$ defining semistrict Lie 2-algebras are equivalent to isomorphism classes of general skeletal semistrict Lie 2-algebras. Moreover, any general semistrict Lie 2-algebras is categorically equivalent to a skeletal one, and therefore the data $(\frg,V,\rho,c)$ can be used to classify semistrict Lie 2-algebras. 

Particularly interesting is the {\em string Lie 2-algebra} of a simple Lie algebra $\frg$, which is defined by the data $(\frg,\FR,\rho,c)$, where $\rho$ is the trivial representation and $c(g_1,g_2,g_3):=k\left<{\rm ad}(g_1),{\rm ad}([g_2,g_3])\right>$, for $k\in\FR$, is a Lie algebra cocycle arising from the Killing form $\left<\cdot,\cdot\right>$ of $\frg$.

\subsection{Differential crossed modules}\label{ssec:diff_crossed_modules}

Both 2- and 3-term $L_\infty$-algebras can be further restricted by demanding that the bracket on $L_0$ satisfies the Jacobi identity (e.g., by putting $\mu_3=0$). For semistrict Lie 2-algebras, this leads to strict Lie 2-algebras, which are equivalent to differential crossed modules \cite{Baez:2002jn}. 

Recall that a \emph{differential crossed module} consists of a pair of Lie algebras $\frg$, $\frh$ together with an action $\acton$ of $\frg$ onto $\frh$ as derivations and a Lie algebra homomorphism between $\frh$ and $\frg$, which we will denote by $\sft$. The maps $\acton$ and $\sft$ satisfy the identities:
\begin{equation}\label{eq:Pfeif}
 \sft(\gamma\acton \chi)=[\gamma,\sft(\chi)]\eand \sft(\chi_1)\acton \chi_2=[\chi_1,\chi_2]
\end{equation}
for all $\gamma\in \frg$ and $\chi,\chi_1,\chi_2\in \frh$. We will denote such a differential crossed module by $\frh\xrightarrow{~\sft~} \frg$.

This structure, when endowed with metrics on $\frg$ and $\frh$, contains the well known metric 3-algebras relevant to M2-brane models \cite{Palmer:2012ya}. Let us briefly recall this relation, focusing on so-called real 3-algebras; hermitian 3-algebras are treated analogously.

A {\em real 3-algebra} \cite{Cherkis:2008qr} is a real vector space $\CA$ endowed with a trilinear map (referred to in the following as a triple bracket) $[\cdot,\cdot,\cdot]:\CA^{\times 3}\rightarrow \CA$, which is antisymmetric in its first two slots, together with an invariant metric. The triple bracket is required to satisfy the so-called {\em fundamental identity}:
\begin{equation}\label{eq:fun}
 [a_1,a_2,[b_1,b_2,b_3]]=[[a_1,a_2,b_1],b_2,b_3]+[b_1,[a_1,a_2,b_2],b_3]+[b_1,b_2,[a_1,a_2,b_3]]
\end{equation}
for all $a_1,a_2,b_1,b_2,b_3\in \CA$. Due to this identity, the span of the operators $D(a,b)$, $a,b\in\CA$, which act on $c\in\CA$ according to
\begin{equation}
 D(a,b)\acton c:=[a,b,c]~,
\end{equation}
forms a Lie algebra, which we will denote by $\frg_\CA$. Endowing $\CA$ with an invariant inner product, we arrive at a {\em real metric 3-algebra}. As shown in \cite{deMedeiros:2008zh}, real metric 3-algebras are in one-to-one correspondence with pairs of Lie algebras $\frg_\CA$ and an orthogonal faithful representation $\CA$ via the Faulkner construction. Explicitly, the metrics on $\frg$ and $\frh$ lead to a triple bracket via 
\begin{equation}\label{eq:metric}
 ( D(a_1,a_2),D(a_3,a_4))_\frg=(D(a_1,a_2)\acton a_3,a_4)_\frh=([a_1,a_2,a_3],a_4)_\frh~.
\end{equation}
Inversely, a triple bracket can be used to define a metric on $\frg$.

Differential crossed modules appear as structure Lie 2-algebras of so-called principal 2-bundles in higher gauge theory. Going one step further in the categorification, principal 3-bundles make use of \emph{differential 2-crossed modules}, which consist of a normal\footnote{Here normal means that the images of the $\sft$ maps are normal subalgebras.} complex of Lie algebras
\begin{equation}
 \frl\ \xrightarrow{~\sft~}\ \frh\ \xrightarrow{~\sft~}\ \frg
\end{equation}
equipped with $\frg$-actions on $\frh$ and $\frl$ by derivations, again denoted by $\acton$, and a $\frg$-equivariant bilinear map, called {\em Peiffer lifting} and denoted by $\{\cdot,\cdot\}: \frh\times \frh\rightarrow \frl$. These maps satisfy the following axioms:
\begin{conditions}
 \item[(i)] $\sft(\gamma\acton \lambda)=\gamma\acton\sft(\lambda)$ and $\sft(\gamma\acton \chi)=[\gamma,\sft(\chi)]$ ,
 \item[(ii)] $\sft(\{\chi_1,\chi_2\})=[\chi_1,\chi_2]-\sft(\chi_1)\acton \chi_2$,
 \item[(iii)] $\{\sft(\lambda_1),\sft(\lambda_2)\}=[\lambda_1,\lambda_2]$,
 \item[(iv)] $\{[\chi_1,\chi_2],\chi_3\}=\sft(\chi_1)\acton\{\chi_2,\chi_3\}+\{\chi_1,[\chi_2,\chi_3]\}-\sft(\chi_2)\acton\{\chi_1,\chi_3\}-\{\chi_2,[\chi_1,\chi_3]\}$,
 \item[(v)] $\{\chi_1,[\chi_2,\chi_3]\}=\{\sft(\{\chi_1,\chi_2\}),\chi_3\}-\{\sft(\{\chi_1,\chi_3\}),\chi_2\}$ ,
 \item[(vi)] $\{\sft(\lambda),\chi\}+\{\chi,\sft(\lambda)\}=-\sft(\chi)\acton \lambda$ ,
\end{conditions}
for all $\gamma\in\frg$, $\chi\in\frh$, and $\lambda\in\frl$, where $[\cdot,\cdot]$ denotes the Lie bracket in the respective Lie algebra. Analogously to the case of a differential crossed module, we denote such a differential 2-crossed module by $\frl\rightarrow \frh\rightarrow \frg$. Note that for trivial $\frl$, a differential 2-crossed module reduces to a differential crossed module. For more details on differential 2-crossed modules, see \cite{Saemann:2013pca} and references therein.

\subsection{(1,0) gauge structures and semistrict Lie 3-algebras}\label{ssec:(1,0)fromLie3}

Consider a (1,0) gauge structure with $\sfg=\sfb=0$. As we saw before in section \ref{ssec:Courant_Dorfman_algebras}, such a (1,0) gauge structure is equivalent to a Courant-Dorfman algebra. It is easy to verify that a Courant-Dorfman algebra $(\CR,\CE,\llbracket\cdot,\cdot\rrbracket)$ gives rise to a semistrict Lie 2-algebra with
\begin{equation}
 L_{-1}=\CR=\frh\eand L_0=\CE=\frg
\end{equation}
as well as higher products
\begin{equation}
\begin{aligned}
 \mu_1(r)&:=\CD r=\sfh(r)~,~~~&\mu_2(e_1,e_2)&:=\llbracket e_1,e_2\rrbracket=-\sff(e_1,e_2)~,\\
 \mu_2(e,r)&:=\tfrac{1}{2}\langle e,\CD r\rangle=\sfd(e,\sfh(r))~,~&\mu_3(e_1,e_2,e_3)&:=-\tfrac{1}{2}\langle e_{[1},\llbracket e_2,e_{3]}\rrbracket\rangle=\sfd(e_{[1},\sff(e_2,e_{3]}))~,
\end{aligned}
\end{equation}
where $e,e_1,e_2,e_3\in\CE$ and $r\in\CR$. In the special case of Courant algebroids, this observation was already made in \cite{Roytenberg:1998vn}.\footnote{As a side remark, note that a Courant-Dorfman algebra with the Dorfman bracket, which is not antisymmetric but satisfies the Jacobi identity, can be regarded as a {\em hemistrict} Lie 2-algebra, cf.\ \cite{Roytenberg:0712.3461}.}

Inversely, many interesting Lie 2-algebras do not form (1,0) gauge structures. For example, consider the Lie 2-algebra based on the octonions with $L_{-1}=L_0=\mathbbm{O}$, where $\mu_2$ is given by the commutator and $\mu_3$ is given by the Jacobiator. In this case, the Jacobiator cannot be written as $\sfd(\cdot,[\cdot,\cdot])$ for any symmetric map $\sfd:\mathbbm{O}\odot\mathbbm{O}\rightarrow\mathbbm{O}$.

For (1,0) gauge structures with $\sfg$ and $\sfb$ nontrivial, the situation is more involved. We evidently start from the chain complex
\begin{equation}\label{eq:chain_complex2}
 L_{-2}=\frg^*\ \xrightarrow{~~\mu_1:=\sfg~~}\ L_{-1}=\frh\ \xrightarrow{~~\mu_1:=\sfh~~}\ L_0=\frg
\end{equation}
together with the maps
\begin{equation}
 \begin{aligned}
 \mu_1(\lambda):=\sfg(\lambda)~,~~~\mu_1(\chi):=\sfh(\chi)\eand \mu_2(\gamma_1,\gamma_2):=-\sff(\gamma_1,\gamma_2)~.
 \end{aligned}
\end{equation}
The higher homotopy relations \eqref{eq:homotopy_relation_a}-\eqref{eq:homotopy_relation_f} then define the remaining products up to terms in the kernels of $\sfg$ and $\sfh$, where the latter turn out to lie in the image of $\sfg$:
\begin{equation}
 \begin{aligned}
  \mu_2(\gamma,\chi)&=\sfd(\gamma,\sfh(\chi))+\sfg(\phi_1(\gamma,\chi))~,\\
  \mu_2(\gamma,\lambda)&=\phi_1(\gamma,\sfg(\lambda))+\phi_2(\gamma,\lambda)~,~~~&\phi_2(\gamma,\lambda)&\in\kernel\,\sfg~,\\
  \mu_2(\chi_1,\chi_2)&=\sfb(\chi_{(1},\sfh(\chi_{2)}))+2\phi_1(\sfh(\chi_{(1}),\chi_{2)})+\phi_3(\chi_1,\chi_2)~,~~~&\phi_3(\chi_1,\chi_2)&\in\kernel\,\sfg~,\\
  \mu_3(\gamma_1,\gamma_2,\gamma_3)&=\sfd(\gamma_{[1},\sff(\gamma_2,\gamma_{3]}))+\sfg(\phi_4(\gamma_1,\gamma_2,\gamma_3))~,\\
  \mu_3(\chi,\gamma_1,\gamma_2)&=-\tfrac{2}{3}\sfb(\sfd(\gamma_{[1},\sfh(\chi)),\gamma_{2]})+2\phi_1(\gamma_{[1},\sfd(\gamma_{2]},\sfh(\chi)))\\
  &\hspace{1.5cm}+2\phi_1(\gamma_{[1},\sfg(\phi_1(\gamma_{2]},\chi)))+\phi_1(\sff(\gamma_1,\gamma_2),\chi)\\
  &\hspace{1.5cm}-\phi_4(\sfh(\chi),\gamma_1,\gamma_2)+\phi_5(\chi,\gamma_1,\gamma_2)~,~~~&\phi_5(\chi,\gamma_1,\gamma_2)&\in\kernel\,\sfg~.
 \end{aligned}
\end{equation}
Equation \eqref{eq:homotopy_relation_i} defines $\mu_4(\gamma_1,\gamma_2,\gamma_3,\gamma_4)$ in a similar way. The challenge is now to fix the $\phi_i$ such that the remaining homotopy relations \eqref{eq:homotopy_relation_g}, \eqref{eq:homotopy_relation_h}, \eqref{eq:homotopy_relation_j} and \eqref{eq:homotopy_relation_k} are satisfied.

A detailed analysis using a computer algebra program suggests that in general, there are no such $\phi_i$ and one has to impose additional constraints onto the (1,0) gauge structure. We understand these constraints as a hint that the (1,0) gauge structure needs to be extended, and there are two possibilities for such extensions. First, the extensions discussed briefly in section \ref{ssec:Bianchi_Extension}, which result in an extended (1,0) gauge structure forming a Lie $n$-algebra with $n>3$. Second, one can extend the chain complex \eqref{eq:chain_complex} to an exact sequence, leading to a Lie 4-algebra. We will discuss this extension briefly in the next section.

But first, let us try to turn the (1,0) gauge structure into a Lie 3-algebra. There is a large number of possible constraints that do this, many of which involve the shifted-graded Jacobi identity for $\sfb$ and $\sfd$ given in equation \eqref{eq:angular_graded_jacobi}. Here we only want to study one. Because we considered the extreme case where $\sfg=0$ (as well as $\sfb=0$) before, let us now turn to the opposite extreme and impose the condition that the kernel of $\sfg$ is trivial. In this case, the maps $\phi_2,\phi_3$ and $\phi_5$ are trivial, and we put
\begin{equation}
 \phi_1(\gamma,\chi):=\alpha_1\sfb(\chi,\gamma)~,~~~\alpha_1\in\FR\eand \phi_4(\gamma_1,\gamma_2,\gamma_3)=0~.
\end{equation}
The map $\mu_4(\gamma_1,\gamma_2,\gamma_3,\gamma_4)$ is given by
\begin{equation}
 \mu_4(\gamma_1,\gamma_2,\gamma_3,\gamma_4)=-2(1+2\alpha_1)\sfb(\sfd(\gamma_{[1},\sff(\gamma_2,\gamma_3)),\gamma_{4]})~.
\end{equation}
If the kernel of $\sfg$ is trivial, these maps satisfy all the homotopy relations \eqref{eq:homotopy_relations} and thus form a semistrict Lie 3-algebra. 

There are two interesting choices for $\alpha_1$. First, the choice  $\alpha_1=-\tfrac{1}{2}$ gives 
\begin{equation}
\mu_2(\gamma,\chi)=\tfrac{1}{2}\rho(\gamma)\acton \chi~,~~\mu_2(\gamma,\lambda)=\tfrac{1}{2}\rho(\gamma)\acton \lambda\eand\mu_4=0~.
\end{equation}

Second, with the choice $\alpha_1=-1$ the curvatures $\CF$ and $\CH$ defined in \eqref{eq:curvatures} can be rewritten in the form 
\begin{equation}
\begin{aligned}
\CF&=\dpar  A+\tfrac{1}{2}\mu_2(A,A)+\mu_1(B)~,\\
\CH&=\dpar  B+ \mu_2(A,B)+\tfrac{1}{6}\mu_3(A,A,A)+\mu_1(C)~,
\end{aligned}
\end{equation}
provided we assume that the {\em fake curvature condition} $\CF=0$ is satisfied. This condition is very natural from the point of view of higher gauge theory, and we will return to it in section \ref{ssec:highergaugetheory}. Note that the Chern-Simons term in $\CH$ collapsed into Lie 3-algebra products. The above form for $\CH$ has been suggested in the context of semistrict higher gauge theory in \cite{Zucchini:2011aa}.

Moreover, demanding that both fake curvatures $\CF$ and $\CH$ vanish and that the graded Jacobi identity \eqref{eq:angular_graded_jacobi} is satisfied, we find that all products in the gauge transformations \eqref{eq:shiftqauge} can be written in terms of Lie 3-algebra products as follows:
\begin{equation}
\begin{aligned}
\delta A=&~\dpar  \alpha+\mu_2(A,\alpha)-\mu_1(\Lambda)~,\\
\delta B=&~\dpar  \Lambda+\mu_2(B,\alpha)+\mu_2(A,\Lambda)+\tfrac{1}{2}\mu_3(A,A,\alpha)-\mu_1(\Xi)~,\\
\delta  C=&~\dpar  \Xi+\mu_2(C,\alpha)+\mu_2(B,\Lambda)+\mu_2(A,\Xi)-\tfrac{1}{2}\mu_3(A,A,\Lambda)+\mu_3(B,A,\alpha)\\&\hspace{1cm}+\tfrac{2}{3}\mu_4(A,A,A,\alpha).
\end{aligned}
\end{equation}
We regard this as a good starting point for studying semistrict higher gauge theory based on Lie 3-algebras. As far as we are aware, this has yet to be developed.

Note however that several terms remain in the supersymmetry transformations and equations of motion which are not of the form of Lie 3-algebra products.

\subsection{Strong homotopy Lie algebras from resolutions of Lie algebras}

Demanding that $\sfg$ is injective is a first step towards turning the chain complex \eqref{eq:chain_complex} underlying the (1,0) gauge structure into an exact sequence. On such sequences, there is a canonical construction of strong homotopy Lie structures \cite{Barnich:1997ij}, as we briefly review in the following. Consider a resolution of a vector space $\frg_0$. That is, consider an exact sequence of vector spaces
\begin{equation}\label{eq:resolution}
 \cdots \xrightarrow{~\mu_1~} L_{-2} \xrightarrow{~\mu_1~} L_{-1} \xrightarrow{~\mu_1~} L_0\xrightarrow{~\mu_1~} \frg_0\xrightarrow{~\mu_1~} 0~.
\end{equation}
Because the sequence is exact, we can decompose $L_0=\frb\oplus \frg_0'$ where $\frb={\rm ker}(\mu_1)$ and $\frg_0'\cong \frg_0$. Assume now that there is a skew-symmetric bilinear map
\begin{equation}
 \mu_2:L_0\times L_0 \rightarrow L_0~,
\end{equation}
which satisfies for all $\ell\in L_0$ and $b\in\frb$ the following two properties:
\begin{conditions}
 \item[(i)] $\mu_2(\ell,b)\in\frb$,
 \item[(ii)] $\mu_2(\mu_2(\ell_1,\ell_2),\ell_3)-\mu_2(\mu_2(\ell_1,\ell_3),\ell_2)+\mu_2(\mu_2(\ell_2,\ell_3),\ell_1)\in \frb$.
\end{conditions}
	  
Then, as shown in \cite{Barnich:1997ij}, the map $\mu_2$ can be extended to a Lie bracket on $\frg_0$ and further to a strong homotopy Lie algebra on all of $L=L_\bullet$. First, one extends $\mu_2$ to all of $L_\bullet$ by showing that
\begin{equation}\label{eq:ext_cond_1}
 \mu_1(\mu_2(\mu_1(\ell_1\otimes \ell_2)))=0~,~~\mbox{for}~ \ell_1,\ell_2\in L_\bullet~.
\end{equation}
As the complex \eqref{eq:resolution} is exact, this equation implies $\mu_2(\mu_1(\ell_1\otimes \ell_2))=\mu_1(\ell_3)$ for some $\ell_3$, and we can define $\mu_2(\ell_1,\ell_2):=\ell_3$. Starting from $\mu_2$ on $L_0\times L_0$, one can iteratively define $\mu_2$ for all higher $L_n$. Note that for $\ell_1,\ell_2\in L_0$, \eqref{eq:ext_cond_1} follows from axiom (i), otherwise one can calculate it using the iteratively defined $\mu_2$. 

For higher products, we use the same method, applied to the corresponding higher Jacobi relations. For example, to define $\mu_3$, we use that
\begin{equation}\label{eq:ext_cond_2}
 \mu_1\big(\mu_3(\mu_1(\ell_1),\ell_2,\ell_3)\pm\mu_2(\mu_2(\ell_2,\ell_3),\ell_1)\pm\mu_2(\mu_2(\ell_1,\ell_2),\ell_3)\pm\mu_2(\mu_2(\ell_1,\ell_3),\ell_2)\big)=0~,
\end{equation}
where the signs are to be chosen according to the gradings of $\ell_1,\ell_2$ and $\ell_3$. Again, for $\ell_1,\ell_2,\ell_3\in L_0$, \eqref{eq:ext_cond_2} follows from axiom (ii), otherwise one can calculate it using the iteratively defined $\mu_3$. Together with the exactness of \eqref{eq:resolution} we thus have
\begin{equation}
 \mu_3(\mu_1(\ell_1),\ell_2,\ell_3)\pm\mu_2(\mu_2(\ell_2,\ell_3),\ell_1)\pm\mu_2(\mu_2(\ell_1,\ell_2),\ell_3)\pm\mu_2(\mu_2(\ell_1,\ell_3),\ell_2)=\mu_1(\ell_4)~,
\end{equation}
for some $\ell_4$, which leads us to define $\mu_3(\ell_1,\ell_2,\ell_3):=\ell_4$.

For a (1,0) gauge structure with $\sfb$ and $\sfg$ trivial, we consider the exact sequence
\begin{equation}\label{eq:chain_complex_3}
 0 \longrightarrow \frh \xrightarrow{~~\sfh~~} \frg \xrightarrow{~~{\rm proj}~~} \frg_0\longrightarrow 0~,
\end{equation}
which induces a splitting $\frg={\rm im}\sfh\oplus \frg_0$. As shown e.g.\ in \cite[sec. 3]{Samtleben:2012mi}, $\frg_0$ forms a Lie algebra with Lie bracket given by $-\sff|_{\frg_0}$. If we now follow the above construction, we recover precisely the Lie 2-algebra structure of a (1,0) gauge structure with $\sfb$ and $\sfg$ trivial: besides $\mu_1(\chi)=\sfh(\chi)$, we have the following higher products:
\begin{equation}
 \mu_2(\gamma_1,\gamma_2)=-\sff(\gamma_1,\gamma_2)~,~~\mu_2(\gamma,\chi)=\sfd(\gamma,\sfh(\chi))~~\mbox{and}~~ \mu_3(\gamma_1,\gamma_2,\gamma_3)=\sfd(g_1,\sff(g_2,g_3))~.
\end{equation}

Assuming that $\sfg$ has trivial kernel and that $\im(\sfg)=\kernel(\sfh)$, we can extend the exact sequence \eqref{eq:chain_complex_3} to 
\begin{equation}\label{eq:chain_complex_3}
 0 \longrightarrow \frg^* \xrightarrow{~~\sfg~~} \frh \xrightarrow{~~\sfh~~} \frg \xrightarrow{~~{\rm proj}~~} \frg_0\longrightarrow 0~.
\end{equation}
The above construction then recovers the Lie 3-algebra that we derived in the previous section with $\alpha_1=0$.

Note that more generally, if $\im(\sfg)=\kernel(\sfh)$, we obtain the exact sequence
\begin{equation}\label{eq:chain_complex_4}
 0 \longrightarrow \ker(\sfg)\longhookrightarrow \frg^* \xrightarrow{~~\sfg~~} \frh \xrightarrow{~~\sfh~~} \frg \xrightarrow{~~{\rm proj}~~} \frg_0\longrightarrow 0~,
\end{equation}
and correspondingly a Lie 4-algebra via the above construction. 

Finally, even if $\im(\sfg)\varsubsetneq\kernel(\sfh)$, we can construct an extension of the map $\sfg:\frg^*\rightarrow \frh$ to a map $\tilde{\sfg}:\frg^*\oplus \fra \rightarrow \frh$ for some vector space $\fra$ such that $\im(\tilde{\sfg})=\kernel(\sfh)$. Then the exact sequence
\begin{equation}\label{eq:chain_complex_4}
 0 \longrightarrow \ker(\tilde{\sfg})\longhookrightarrow \frg^*\oplus\fra \xrightarrow{~~\tilde{\sfg}~~} \frh \xrightarrow{~~\sfh~~} \frg \xrightarrow{~~{\rm proj}~~} \frg_0\longrightarrow 0
\end{equation}
yields again a Lie 4-algebra.

Since higher gauge theory has not been developed for Lie 4-algebras, our subsequent discussion has to remain restricted to (1,0) gauge structures that form Lie 3-algebras.

\section{Examples}

Above, we have seen that (1,0) gauge structures contain semistrict Lie 2-, Lie 3- or Lie 4-algebras, which in turn suggests that the new (1,0) models may be formulated in terms of higher gauge theory. In this section, we will briefly review some notions of higher gauge theory, before we then discuss various interesting examples of (1,0) gauge structures.

\subsection{Higher gauge theory}\label{ssec:highergaugetheory}

Higher gauge theory is the theory of parallel transport of extended objects. In particular, it makes use of categorified versions of principal fiber bundles where categorified versions of Lie groups take over the role of the structure groups. To the detail necessary for our discussion, higher gauge theory has only been developed for principal 2- and 3-bundles having differential crossed and 2-crossed modules as their higher structure Lie algebras, respectively. For a discussion of the case of principal 2-bundles with connective structure, see \cite{Baez:2002jn,Baez:2004in,Baez:2010ya}, for the case of principal 3-bundles, see \cite{Martins:2009aa} and in particular \cite{Saemann:2013pca}.

As the (1,0) model is a local theory, we can restrict ourselves to the local description in terms of gauge potentials. In the case of principal 2-bundles with connective structure, we have a differential crossed module $\frh\xrightarrow{~\sft~} \frg$ as gauge Lie 2-algebra and potential 1- and 2-forms, $A$ and $B$, taking values in the Lie algebras $\frg$ and $\frh$, respectively. Their curvatures read as
\begin{equation}\label{eq:DefOfFH}
 F := \dpar  A+\tfrac{1}{2}[A, A]\eand H:=\nabla B:=\dpar  B+A\acton B~.
\end{equation}
It has been shown in several contexts, cf.\ \cite{Baez:2004in,Baez:2010ya}, that for these curvatures to describe a consistent parallel transport of an extended object, it is crucial that 
the {\em fake curvature condition}
\begin{equation}\label{eq:fake_curvature}
 \CF:=\ F + \sft(B)=0
\end{equation}
is imposed. Otherwise, the parallel transport will {\em not} be invariant under worldsheet reparameterizations of the extended object.

Infinitesimal gauge transformations are parametrized by a $\frg$-valued function $\alpha$ as well as a one-form $\Lambda$ taking values in $\frh$:
\begin{equation}\label{eq:Space-time-GT}
\begin{aligned}
\delta A&=\dpar  \alpha+[A,\alpha]-\sft(\Lambda)~,\\
\delta B&=\dpar  \Lambda +A\acton \Lambda-\alpha\acton B~.
\end{aligned}
\end{equation}
In particular, the fake curvature condition \eqref{eq:fake_curvature} is invariant under gauge transformations. Note that the curvature \eqref{eq:curvatureF10} of the (1,0) model has to be identified with the fake curvature $\CF$, as it is the only two-form curvature built from $A$ and $B$ that transforms covariantly.

In the case of principal 3-bundles, we have a differential 2-crossed module $\frl\xrightarrow{~\sft~}\frh\xrightarrow{~\sft~} \frg$ together with potential 1-, 2- and 3-forms $A$, $B$ and $C$, which take values in the Lie algebras $\frg$, $\frh$ and $\frl$, respectively. Their curvatures are given by
\begin{equation}
 F\ :=\ \dpar  A+\tfrac{1}{2}[A,A]~,~~H\:=\ \dpar  B+A\acton B~,~~G\ :=\ \dpar  C+A\acton C+\{B,B\}~,
\end{equation}
and there are two fake curvature conditions:
\begin{equation}\label{eq:fake_curvature_3}
 \CF :=F+\sft(B)=0\eand \CH:=H+\sft(C)=0~.
\end{equation}
Gauge transformations \cite{Saemann:2013pca} now involve an additional two-form parameter $\Xi\in\frl$ :
\begin{equation}\label{eq:Space-time-GT2}
\begin{aligned}
\delta A&=\dpar  \alpha+[A,\alpha]-\sft(\Lambda)~,\\
\delta B&=\dpar  \Lambda+A\acton \Lambda-\alpha\acton B -\sft(\Xi)~,\\
\delta C&=\dpar \Xi+A\acton\Xi-\alpha\acton C-\{B,\Lambda\}-\{\Lambda,B\}~.
\end{aligned}
\end{equation}
These gauge transformations \eqref{eq:Space-time-GT2} and fake curvature conditions \eqref{eq:fake_curvature_3} arose naturally in a twistor construction of (2,0) superconformal field configurations in \cite{Saemann:2013pca}, along with equations of motion, which, in a certain gauge, include $H=*H$. 

Let us stress here that the fake curvature condition $\CF=0$ is {\em not} stable under supersymmetry transformations \eqref{eq:susy} in general. Therefore, whenever we impose the fake curvature condition in the following, we implicitly break supersymmetry. A way out of this problem would be to impose, in addition, the equations arising from a supersymmetry variation of the fake curvature condition, as well as further equations arising from supersymmetry variations of the latter.

Note that in the models arising from the above mentioned twistor constructions, the fake curvature condition is indeed invariant under the corresponding supersymmetry transformations. 

\subsection{Abelian gerbe} 

Our first example is the simplest, that of an abelian gerbe, cf.\ \cite{Hitchin:1999fh}. If we take the vector spaces 
\begin{equation}
0\longrightarrow\fru(1)\longrightarrow0~, 
\end{equation}
and set all the maps to zero, we are left with just the (1,0) tensor multiplet $(\phi,\chi,B)$ satisfying the equations of motion
\begin{equation}
\CH=\dpar  B=*\CH~,~~\slasha{\dpar}\chi=0\eand\Box \phi=0~,
\end{equation}
and transforming under the usual gauge transformation for an abelian gerbe 
\begin{equation}
\delta B=\dpar  \Lambda~.
\end{equation}

The supersymmetry transformations become
\begin{equation}\label{eq:susy2}
\begin{aligned}
\delta \phi&=\epsb\chi~,~~~
\delta \chi^i&=\tfrac{1}{8}\slasha{\CH}\eps^i+\tfrac{1}{4}\slasha{\dpar}\phi~\eps^i~,~~~
\delta B&=-\epsb\gamma^{(2)}\chi~,
\end{aligned}
\end{equation}
which match the full $(2,0)$ supersymmetry transformations for a single M5-brane \cite{Howe:1997fb} when reduced to a contained (1,0) multiplet.

\subsection{Principal 2-bundles}

For non-abelian gerbes, we use the language of principal 2-bundles and differential crossed modules.

To obtain differential crossed modules from a (1,0) gauge structure, we set $\sfg=\sfb=0$ and assume 
\begin{equation}\label{eq:Jacobi}
\begin{aligned}
\sfd(\sff(\gamma_{[1},\gamma_2),\gamma_{3]})=0~.
\end{aligned}
\end{equation}
This ensures that $\sff(\cdot,\cdot)$ is a Lie bracket on $\frg$ by \eqref{eq:algebra_relation_e} and corresponds to setting $\mu_3=0$ on the Lie 2-algebra level, making it a strict Lie 2-algebra. Nontrivial such (1,0) gauge structures are very restricted, but can indeed be constructed, e.g.\ by using the analysis in \cite[sec.\ 3]{Samtleben:2012mi}.

Now to obtain a differential crossed module we define 
\begin{equation}
\begin{aligned}
\sft:=\sfh~,~~~~[\gamma_1,\gamma_2]:=-\sff(\gamma_1,\gamma_2)\eand\gamma\acton\chi:=\sfd(\gamma,\sfh(\chi))~.
\end{aligned}
\end{equation}
Note that this is a differential crossed module with abelian $\frh$ since $[\chi_1,\chi_2]=\sft(\chi_{[1})\acton \chi_{2]}=\sfd(\sfh(\chi_{[1}),\sfh(\chi_{2]}))=0$, by \eqref{eq:Pfeif} and the symmetry of $\sfd$.

Note also that for $\sfg=0$ the vector multiplet equations of motion \eqref{eq:eom2} become trivial, and we can therefore eliminate the degrees of freedom by enforcing the fake curvature condition \eqref{eq:fake_curvature} of higher gauge theory:
\begin{equation}
\begin{aligned}
\CF=\dpar  A-\tfrac{1}{2}\sff(A,A)+\sfh(B)=0~.
\end{aligned}
\end{equation}
Using \eqref{eq:Jacobi}, the shifted form of the (1,0) gauge transformations \eqref{eq:shiftqauge} becomes
\begin{equation}
\begin{aligned}
\delta A&=\dpar  \alpha-\sff(A,\alpha)-\sfh(\Lambda)~,\\
\delta B&=\dpar  \Lambda+\sfd(A,\sfh(\Lambda))-\sfd(\alpha,\sfh(B))~,
\end{aligned}
\end{equation}
which matches exactly the higher gauge theory gauge transformations \eqref{eq:Space-time-GT}.

One of the most interesting classes of differential crossed modules is that of the 3-algebras appearing in the context of M2-brane models, cf.\ section \ref{ssec:diff_crossed_modules}. However these are not included in the above discussion since they have a trivial map $\sft=0$ and a non trivial action $\acton$. Since the maps above were defined by $\gamma\acton\chi:=-\sfd(\gamma,\sfh(\chi))$ and $\sft:=\sfh$, a trivial map $\sft$ implies a trivial action. Luckily 3-algebras can be treated separately, and we will come back to them shortly.

\subsection{Principal 3-bundles}

Higher gauge theory has been developed not only for principal 2-bundles but also for principal 3-bundles, which have differential 2-crossed modules as underlying structure Lie 3-algebras. For this section we assume first that the products corresponding to Lie 3-algebra products $\mu_3$ and $\mu_4$ are zero:  
\begin{equation}\label{eq:Jacobi2}
 \begin{aligned}
\sfd(\gamma_{[1},\sff(\gamma_2,\gamma_{3]}))=0~,~~~\sfb(\sfd(\gamma_{[1},\sff(\gamma_2,\gamma_3)),\gamma_{4]})=0~&,\\
\sfb(\chi,\sff(\gamma_1,\gamma_2))-\tfrac{4}{3}\sfb(\sfd(\gamma_{[1},\sfh(\chi)),\gamma_{2]})+2\sfb(\sfg(\sfb(\gamma_{[1},\chi)),\gamma_{2]})&=0~,\\
 \end{aligned}
\end{equation}
and second that terms of the form $\sfb(\sfg(\cdot),\sfh(\cdot))$ vanish. These terms are in the kernel of $\sfg$ and are therefore expected to vanish, as discussed in section \ref{ssec:(1,0)fromLie3}. 

To obtain differential 2-crossed modules from (1,0) gauge structures we define
\begin{equation}
\begin{aligned}
\sft&(\lambda):=\sfg(\lambda)~,~~\sft(\chi):=\sfh(\chi)~,~~[\gamma_1,\gamma_2]:=-\sff(\gamma_1,\gamma_2)~,~~\gamma\acton\chi:=\sfd(\gamma,\sfh(\chi))-\sfg(\sfb(\chi,\gamma))~,\\
[&\chi_1,\chi_2]=[\lambda_1,\lambda_2]=0~,~~\{\chi_1,\chi_2\}:=\tfrac{1}{2}\sfb(\chi_{1},\sfh(\chi_{2}))\eand \gamma\acton\lambda:=-\sfb(\sfg(\lambda),\gamma)~.
\end{aligned}
\end{equation}
Note that this is a differential 2-crossed module with abelian $\frl$ and $\frh$.

To reduce to principal 3-bundles, we have to impose the vanishing of the fake curvatures
\begin{equation}
\begin{aligned}
\CF&=\dpar  A-\tfrac{1}{2}\sff(A,A)+\sfh(B)=0~,\\
\CH&=\dpar  B+ 2\sfd(A,\sfh(B))-\sfg(\sfb(B,A))+\sfd(A,\dpar  A-\tfrac{1}{3}\sff(A,A))+\sfg(C)\\&=\dpar  B+ \sfd(A,\sfh(B))-\sfg(\sfb(B,A))+\sfg(C)=0~.
\end{aligned}
\end{equation}

This simplifies the shifted gauge transformations \eqref{eq:shiftqauge} to
\begin{equation}
\begin{aligned}
\delta A&=\dpar  \alpha-\sff(A,\alpha)-\sfh(\Lambda)~,\\
\delta B&=\dpar  \Lambda+\sfd(A,\sfh(\Lambda))+\sfg(\sfb(\Lambda,A))-\sfd(\alpha,\sfh(B))+\sfg(\sfb(B,\alpha))-\sfg(\Xi)~,\\
\delta  C&=\dpar  \Xi-\sfb(\sfg(\Xi),A)+\sfb(\sfg(C),\alpha)-\sfb(B,\sfh(\Lambda)) +\dots~,
\end{aligned}
\end{equation}
which match exactly the higher gauge theory transformations \eqref{eq:Space-time-GT2}.

The constraints \eqref{eq:Jacobi2} are again very restrictive. One admissible example is the Chern-Simons Lie 3-algebra of $\fru(1)$, which we will discuss in section \ref{ssec:chern}. If we are just interested in the algebraic structure and not in matching the gauge transformations to higher gauge theory, we can discuss many more interesting examples. 

\subsection{Representations of Lie algebras and M2-brane model 3-algebras}

Let $\fra$ be a semi-simple Lie algebra with a representation $\rho$ acting on a vector space $V$. There are three types of models based on this information, as discussed in \cite{Samtleben:2012mi}; here we will just discuss the simplest one. An action is not possible for this type, however the type admitting an action is closely related.

We take the complex
\begin{equation}
0\longrightarrow V\longrightarrow V\times\fra~,
\end{equation}
and choose the maps 
\begin{equation}
 \begin{aligned}
\sfg=\sfb=&0~,~~\sfh(v)=\binom{v}{ 0}~,\\
\sfd\left(\binom{v_1}{ g_1},\binom{v_2}{ g_2}\right)&=\tfrac{1}{2}(\rho(g_1)\acton v_2+\rho(g_2)\acton v_1)~,~~\\
\sff\left(\binom{v_1}{ g_1},\binom{v_2}{ g_2}\right)&=\binom{\tfrac{1}{2}(\rho(g_2)\acton v_1-\rho(g_1)\acton v_2)}{[g_1,g_2]}~,
 \end{aligned}
\end{equation}
for $v\in V,~\binom{v_i}{g_i}\in V\times\fra$.

Recall that metric 3-algebras are obtained from metric Lie algebras with faithful orthogonal representations via the Faulkner construction \cite{deMedeiros:2008zh}, where the representation space is  the 3-algebra itself, $V=\CA$, and the Lie algebra is the associated Lie algebra of inner derivations 
 $\fra=\frg_\CA$. 

In order to use this relation we need to endow the (1,0) gauge structure with metrics on the spaces $\fra$ and $V$ which are invariant under the action of $\fra$. Explicitly, this construction gives the triple bracket
\begin{equation}\label{eq:triple}
[v_1,v_2,v_3]=\sfd(\sfm^*_\fra( \sfd^*(\sfm_\frh(v_1),\sfh(v_2))),\sfh(v_3))~,
\end{equation}
where $\sfm^*_\fra:\fra^*\rightarrow\fra$  and $\sfm_\frh:\frh\rightarrow\frh^*$ are maps induced from the metrics on $\fra$ and $\frh$, respectively. 

Note that here we constructed a triple bracket using a metric on $\fra$. Inversely, one can derive a metric on $\fra$ given a triple bracket as done in \eqref{eq:metric}.

The simplest non-trivial example is that of $A_4$, which appears in the description of two M2-branes in the BLG model. We choose the fundamental representation of $\fra=\aso(4)$ acting on $V=\FR^4$, along with the standard euclidean metric on $\FR^4$ and a split signature metric on $\aso(4)$, explicitly:\begin{equation}
\begin{aligned}
\sfm_{\aso(4)}(A^{\pm})=\pm \big(A^{\pm}\big)^T~&,~~\sfm_{\FR^4}(v)=v^T~,\\
\sfd\left(\binom{v}{A},\binom{w}{B}\right)=\tfrac{1}{2}(A.w+B.v)~&,~~\sfd^*\left(v^T,\binom{w}{A}\right)=\tfrac{1}{2}\binom{v^T. A}{w v^T- vw^T}~,
\end{aligned}
\end{equation}
for $v,w\in \FR^4,~A,B\in\aso(4)$ and where $v^T$ denotes the transpose of $v$ and $A^\pm$ denote the selfdual and anti-selfdual parts of $A$. Then \eqref{eq:triple} gives the triple bracket on the basis vectors $e^\mu\in\FR^4$ as
\begin{equation}
[e^\mu,e^\nu,e^\rho]=\eps^{\mu\nu\rho\sigma}e^\sigma~.
\end{equation}

Similarly, the 3-algebra describing $N$ M2-branes in the ABJM model corresponds to the choice $\fra=\fru(N)\times\fru(N)$, with split signature metric
\begin{equation}
\sfm^*_\fra\binom{A_L}{A_R}=\binom{A_L^\dagger}{-A_R^\dagger}~,
\end{equation}
and where $V=\au(N)$ is the bi-fundamental representation with the standard Hilbert-Schmidt metric $\sfm_\frh(A)=A^\dagger$. The triple bracket then becomes
\begin{equation}
[A,B;C]=\sfd(\sfm^*_\fra( \sfd^*(\sfm_\frh(A),\sfh(C))),\sfh(B))=AC^\dagger B-B C^\dagger A~.
\end{equation}
For $N=2$ this essentially coincides with the 3-Lie algebra $A_4$. 

We can now rewrite equations \eqref{eq:eom} in terms of the products appearing in 3-algebras. Note however a crucial difference here to the M2-brane models: the gauge field of M2-brane models lives only in $\fra$ and not in $V\times \fra$ and also that the gauge transformations have only one ($\fra$-valued) parameter. 

\subsection{Vectors in $\sG\times \sG$} 

Another example is found in \cite{Chu:2011fd}, where the $\sG\times \sG$-model is conjectured to describe the gauge sector of M5-brane dynamics. This conjecture passes many consistency checks, including self-dual string profiles which match gravity dual predictions \cite{Chu:2013hja}. One key difference between the $\sG\times \sG$-model and the (1,0) model is that in the former, the vector fields are on shell and that they are related to the tensor fields in a way reminiscent of the fake curvature condition \eqref{eq:fake_curvature}. Nevertheless, the algebraic structure is an example of a $(1,0)$ gauge structure with matching gauge transformations. In our notation, the vector spaces present are
\begin{equation}
0\longrightarrow\frg\longrightarrow\frg\times\frg~,
\end{equation}
where $\frg$ is a Lie algebra with Lie bracket $[\cdot,\cdot]$. We will use the notation $A=\binom{A_L}{A_R}$ and $\alpha=\binom{\alpha_L}{\alpha_R}$ to denote one-forms and functions taking values in $\frg\times\frg$. The gauge transformations take the following form:
\begin{equation}\label{eq:chu}
 \begin{aligned}
\delta A&=\dpar  \alpha+\binom{[A_L,\alpha_L+\alpha_R]}{[A_R,\alpha_L+\alpha_R]}+\binom{\Lambda}{-\Lambda}~,\\
\delta B&=\dpar  \Lambda+\tfrac{1}{2}[A_L+A_R,\Lambda]+\tfrac{1}{2}([A_R,\dpar \alpha_L]-[A_L,\dpar \alpha_R])+[B,\alpha_L+\alpha_R]~,
 \end{aligned}
\end{equation}
where, as before, wedge products are implied, e.g. $[A_L,\Lambda]=[A_L{}_{\mu},\Lambda_{\nu}]\dd x^\mu\wedge\dd x^\nu$ .

To make contact with the (1,0) gauge structure transformations \eqref{eq:qauge} we set $\sfg=\sfb=0$ and introduce the new shift of gauge parameters $(\alpha,\Lambda)\rightarrow(\alpha,\Lambda+2\sfd(\alpha,A))$. Using \eqref{eq:algebra_relation_a}, we obtain
\begin{equation}\label{eq:chu2}
\begin{aligned}
\delta A&=\dpar  \alpha -\sff(A,\alpha)-\sfh(\sfd(A,\alpha))-\sfh(\Lambda)~,\\
\delta B&=\dpar  \Lambda+\sfd(A,\sfh(\Lambda)-\dpar  \alpha)-2\sfd(\alpha,\sfh(B))~.
\end{aligned}
\end{equation}

With the following choice of maps\footnote{A slightly different set of maps, which satisfy the constraints \eqref{eq:algebra_relations}, was given in \cite{Samtleben:2011fj}. These, however, do not lead to the gauge transformations of \cite{Chu:2011fd}.}
\begin{equation}
 \begin{aligned}
\sfh(g)&=\binom{-g}{ g}~,~~\sfd\left(\binom{g_1}{ g_2},\binom{g_3}{ g_4}\right)=\tfrac{1}{2}([g_1,g_4]+[g_3,g_2])~,~~\\&\sff\left(\binom{g_1}{ g_2},\binom{g_3}{ g_4}\right)=\binom{-[g_1,g_3]-\tfrac{1}{2}([g_1,g_4]-[g_3,g_2])}{-[g_2,g_4]-\tfrac{1}{2}([g_1,g_4]-[g_3,g_2])}~,
 \end{aligned}
\end{equation}
the shifted gauge transformations \eqref{eq:chu2} match \eqref{eq:chu} exactly.

Since this $\sff$ does not satisfy the Jacobi identity, this is not a differential crossed module. However since the above (1,0) gauge structure has trivial maps $\sfg$ and $\sfb$, it is an example of a (semistrict) Lie 2-algebra.

\subsection{String Lie 2-algebras}

Another interesting Lie 2-algebra related to M-theory dynamics is the string Lie 2-algebra \cite{Baez:2003aa}, defined in section \ref{ssec:Lie_2_algebras}. The Lie algebra $\frg$ is put into the complex

\begin{equation}
0\longrightarrow\FR\longrightarrow\frg~,
\end{equation}
and the Lie bracket and Killing form $\left<\cdot,\cdot\right>$ correspond to the maps
\begin{equation}
\begin{aligned}
 \sfg=\sfb=\sfh=0~,~~\sff(\gamma_1,\gamma_2):=-[\gamma_1,\gamma_2]\eand\sfd(\gamma_1,\gamma_2)
=\left<\gamma_1,\gamma_2\right>~.
\end{aligned}
\end{equation}

This model describes an abelian tensor multiplet sourced by a non-abelian vector multiplet. It was originally found in \cite{Bergshoeff:1996qm} and it provided crucial inspiration for the development of the (1,0) superconformal models of \cite{Samtleben:2011fj}. The equations of motion \eqref{eq:eom} now read as
\begin{equation}
\begin{aligned}
\CH^-&=-\left<\lambdab,\gamma^{(3)}\lambda\right>~,\\
\slasha{\dpar} \chi^i&=\left<\slasha{\CF},\lambda^i\right>+2\left<Y^{ij},\lambda_j\right>~,\\
\dpar^2 \phi&=2\left<Y^{ij},Y_{ij}\right>-*2\left<\CF,*\CF\right>-4\left<\lambdab,\slasha{\dpar} \lambda\right>~,
\end{aligned}
\end{equation}
where the field strengths are
\begin{equation}
\begin{aligned}
\CF&=&\dpar  A+\tfrac{1}{2}[A,A]\eand
\CH&=&\dpar  B+\left<A,\dpar  A+\tfrac{1}{3}[A,A]\right>~.
\end{aligned}
\end{equation}
The gauge and supersymmetry transformations can be easily read off from \eqref{eq:qauge} and \eqref{eq:susy}.

\subsection{Chern-Simons Lie 3-algebra}\label{ssec:chern}

In the Chern-Simons Lie 3-algebra $\frcs_k(\frg)$ of a simple Lie algebra $\frg$, the map $\mu_1:L_{-1}\rightarrow L_0$ is surjective. This map should be identified with the map $\sfh$ in a (1,0) gauge structure, and because of \eqref{eq:algebra_relation_d}, this implies that $\sff=0$. We therefore have to restrict ourselves to abelian $\frg$. The Chern-Simons Lie 3-algebra $\frcs_k(\FR)$ consists of the complex 
\begin{equation}\label{eq:chain_complex_chern_simons_3_algebra}
\FR\longrightarrow\FR\times\FR\longrightarrow\FR~,
\end{equation}
with the following non trivial products
\begin{equation}
  \mu_2\left(\gamma_1,\binom{\lambda}{\gamma_2}\right):=\binom{k\gamma_1\gamma_2}{0}\eand
  \mu_2\left(\binom{\lambda_1}{\gamma_1},\binom{\lambda_2}{\gamma_2}\right):=2k\gamma_1\gamma_2~.
\end{equation}
Note that the chain complex \eqref{eq:chain_complex_chern_simons_3_algebra} forms an exact sequence. By the identification of (1,0) gauge structures and Lie 3-algebras based on exact sequences we set
\begin{equation}
\begin{aligned}
 \sfg(\lambda):=\binom{\lambda}{0}~,~~~&\sfh\binom{\lambda}{\gamma}:=\gamma~,~~~\sff=0~,\\
 \sfd(\gamma_1,\gamma_2):=\binom{k\gamma_1\gamma_2}{0}~,~~~&\sfb\left(\binom{\lambda_1}{\gamma_1},\gamma_2\right):=2k\gamma_1\gamma_2~.
\end{aligned}
\end{equation}

The field strengths of $A$ and $B=\binom{B_L}{B_R}$ then read explicitly as
\begin{equation}
\begin{aligned}
\CF&=&\dpar  A+B_R\eand
\CH&=&\dpar  B+\binom{kA\wedge\dpar  A+C}{0}~.
\end{aligned}
\end{equation}
The gauge and supersymmetry transformations become
\begin{equation}
\begin{aligned}
\delta A&=\dpar  \alpha -\Lambda_R~,\\
\delta B&=\dpar  \Lambda+\binom{k A\wedge(\dpar  \alpha -\Lambda_R)-2k\alpha\CF-\Xi}{0}~,\\
\delta  C&=\dpar  \Xi+2k(\dpar  \alpha\wedge B_R+\Lambda_R\wedge\dpar  A+\alpha\CH)~,
\end{aligned}
\end{equation}
and
\begin{equation}\nonumber
\begin{aligned}
\delta \phi&=\epsb\chi~,~~~&\delta  Y^{ij}&=-\epsb^{(i}\slasha{\dpar} \lambda^{j)}+2 \epsb^{(i}\chi^{j)}_R~,\\
\delta \chi^i&=\tfrac{1}{8}\slasha{\CH}\eps^i+\tfrac{1}{4}\slasha{\dpar}\phi~\eps^i-\tfrac{k}{2}\binom{*(\gamma\lambda^i\wedge*\epsb\gamma\lambda)}{0}~,~~~&\delta \lambda^i&=\tfrac{1}{4}\slasha{\CF}\eps^i-\tfrac{1}{2}Y^{ij}\eps_j+\tfrac{1}{4}\phi_R\eps^i~,\\
\delta B&=-k\binom{A\wedge\epsb \gamma \lambda}{0}-\epsb\gamma^{(2)}\chi~,~~~&\delta A&=-\epsb \gamma \lambda~,
\end{aligned}
\end{equation}
\begin{equation}
 \delta C~=~-2k~(~B_R\wedge\epsb \gamma\lambda+\phi_R\epsb\gamma^{(3)}\lambda~)~,
\end{equation}
while the equations of motion read as
\begin{equation}
\begin{aligned}
\CH^-&=-\binom{k\lambdab\gamma^{(3)}\lambda}{0}~,\\
\slasha{\dpar} \chi^i&=k\binom{\slasha{\CF}\lambda^i+2Y^{ij}\lambda_j-3\phi_R\lambda^i}{0}~,\\
D^2 \phi&=2k\binom{Y^{ij}Y_{ij}-*(\CF\wedge*\CF)-2\lambdab\slasha{\dpar} \lambda+2\chib_R\lambda+8\lambdab\chi_R-\tfrac{3}{2}\phi_R^2}{0}~,
\end{aligned}
\end{equation}
\begin{equation}
\begin{aligned}
\phi_R Y^{ij}+2\chib_R^{(i}\lambda^{j)}&=0~,\\
4k(\phi_R\CF+2\chib_R\gamma^{(2)}\lambda)&=*\CH^{(4)}~,\\
\phi_R\slasha{\dpar} \lambda_i+\tfrac{1}{2}\slasha{\dpar}\phi_R \lambda_i&=\tfrac{1}{2}\slasha{\CF}\chi_{Ri}+\tfrac{1}{4}\slasha{\CH}_R\lambda_i-\chi_R^jY_{ij}+\tfrac{3}{2}\phi_R\chi_R~,
\end{aligned}
\end{equation}
where we used the notiation $\phi=\binom{\phi_L}{\phi_R}$ for fields $\phi\in\FR\times \FR$. Note that the field equations all remain interacting.

\subsection{Extreme Courant-Dorfman algebras}

Finally, let us briefly comment on the example of extreme Courant-Dorfman algebras with either $\sfh=0$ or $\sfd=0$ for which $\frg$ is a Lie algebra. In the first case, $\frg$ is a Lie algebra endowed with an invariant quadratic form over $\frh$. Here, we obtain a free (1,0) vector multiplet together with a tensor multiplet in the background of this vector multiplet. Furthermore, the tensor multiplet fields do not interact among each other; all interactions arise from source terms containing exclusively fields of the vector multiplet.

In the second case $\sfd=0$, $\frg$ is a Lie algebra over $\frh$ and $\sfh$ is a derivation with values in the center of $\frg$. The definitions of $\CF$ and $\CH$ correspond to the fake curvature and the curvature 3-form of a principal 2-bundle with strict structure 2-group. The action of the covariant derivative becomes trivial on $\frh$, and we obtain an abelian free tensor multiplet together with a free vector multiplet.

\section*{Acknowledgements}
We would like to thank Richard Szabo and Martin Wolf for discussions. This work was supported by an EPSRC Career Acceleration Fellowship.

\appendices

\subsection{Strong homotopy Lie algebras}\label{app:SH_Lie_algebras}

Recall that a {\em strong homotopy Lie algebra} or {\em $L_\infty$-algebra} is a graded vector space $L=\oplus_n L_n$, equipped with graded antisymmetric multilinear maps
\begin{equation}
 \mu_i:L^{\wedge i}\rightarrow L~,~~~i\geq 1~,
\end{equation}
of degree $2-i$, such that the following higher Jacobi relations are satisfied for each\footnote{Sometimes, a zero-bracket is introduced in addition and $L_\infty$-algebras for which this bracket vanishes (as in our definition) are called `strict'. This nomenclature unfortunately collides with that of a strict $n$-category and we will not use it here.} $m\geq 1$ and homogeneous elements $\ell_1,\ldots,\ell_m$:
\begin{equation}\label{eq:homotopyJacobi}
 \sum_{i+j=m}\sum_\sigma\chi(\sigma;\ell_1,\ldots,\ell_m)(-1)^{i\cdot j}\mu_{j+1}(\mu_i(\ell_{\sigma(1)},\cdots,\ell_{\sigma(i)}),\ell_{\sigma(i+1)},\cdots,\ell_{\sigma(m)})=0~.
\end{equation}
Here, the sum over $\sigma$ is taken over all $(i,j)$ unshuffles. Recall that a permutation $\sigma$ of $i+j$ elements is called an {\em $(i,j)$-unshuffle}, if the first $i$ and the last $j$ images of $\sigma$ are ordered: $\sigma(1)<\cdots<\sigma(i)$ and $\sigma(i+1)<\cdots<\sigma(i+j)$. Moreover, $\chi(\sigma;\ell_1,\ldots,\ell_n)$ is the skew-symmetric Koszul sign defined implicitly via
\begin{equation}
 \ell_1\wedge \ldots \wedge \ell_m=\chi(\sigma;\ell_1,\ldots,\ell_m)\ell_{\sigma(1)}\wedge \ldots \wedge\ell_{\sigma(m)}~,
\end{equation}
where $\wedge$ is seen as a graded anticommutative operation. 

In this paper, we will only be interested in $L_\infty$-algebras which consist of graded vector spaces with non-positive gradings. If the degrees of the vector spaces $L_n$ are further truncated and the $L_\infty$-algebra is concentrated in degrees $-n+1$ to $0$, we call the resulting $L_\infty$-algebra a (semistrict\footnote{General (or {\em weak}) Lie $n$-algebras arise as categorifications of the notion of a Lie algebra, see e.g.\ \cite{Roytenberg:0712.3461}. In this paper, however, we only needed semistrict Lie $n$-algebras.}) {\em Lie $n$-algebra} or {\em $L_n$-algebra}.  

There is an elegant alternative definition of an $L_\infty$-algebra that makes use of a nilpotent differential. First, note that if we shift the grading of an $L_\infty$-algebra $L$ by $-1$ and consider $L[-1]=\oplus_n L_n[-1]$, where $L_n[-1]$ has now grading $n-1$, the degree of all brackets $\mu_i$ becomes +1. After the shift, we can define an $L_\infty$-algebra as a $\RZ^{<0}$-graded vector space $L$ equipped with a differential $\CD:\wedge^\bullet L\rightarrow \wedge^\bullet L$ of degree 1, which satisfies $\CD^2=0$. The connection to the previous definition is made by decomposing
\begin{equation}
 \CD=\CD_1+\CD_2+\CD_3+\cdots
\end{equation}
and demanding that $\CD_i$ acts on elements of $\wedge^i L$ as $\mu_i$, and otherwise it is extended to a coderivation via
\begin{equation}
 \mu_i(\ell_1\wedge\cdots\wedge \ell_m)=\sum_\sigma \chi(\sigma;\ell_1,\ldots,\ell_m)(-1)^{i\cdot (m-i)} \mu_i(\ell_{\sigma(1)},\cdots,\ell_{\sigma(i)})\wedge \ell_{\sigma(i+1)}\wedge \cdots\wedge \ell_{\sigma(m)}~.
\end{equation}
From here, it is rather obvious that the condition $\CD^2=0$ on $\wedge^\bullet L$ translates into the higher Jacobi relations \eqref{eq:homotopyJacobi}.

If all the homogeneously graded vector subspaces $L_n$ of $L$ are finite-dimensional, we can dualize this construction and obtain the {\em Chevalley-Eilenberg algebra} $\CEa(L)=(\wedge^\bullet L^*,Q)$ of $L$, where $Q$ is the dual of $\CD$. The Chevalley-Eilenberg algebra can be regarded as the polynomials on the space $L[-1]$ and $Q:\CEa(L)\rightarrow \CEa(L)$ becomes a homological vector field of degree 1. Altogether, we thus reinterpreted an $L_\infty$-algebra in terms of a {\em $Q$-manifold} as defined in \cite{Alexandrov:1995kv}.


\end{document}